\documentclass{vldb_obfuscated}
\usepackage{booktabs}
\usepackage[table]{xcolor}
\usepackage{makecell}
\usepackage{graphicx}
\usepackage{amssymb}
\usepackage{ifsym}
\usepackage[colorlinks]{hyperref}
\usepackage{balance}  
\usepackage{boxedminipage2e}
\usepackage{subcaption}
\usepackage{array}
\usepackage{algorithm}
\usepackage[noend]{algpseudocode}
\usepackage{listings}
\usepackage{color}
\usepackage{mdwlist}

\definecolor{deepblue}{rgb}{0,0,0.5}
\definecolor{deepred}{rgb}{0.6,0,0}
\definecolor{deepgreen}{rgb}{0,0.5,0}

\definecolor{drawiodarkblue}{HTML}{7BABE0}
\definecolor{drawiodarkred}{HTML}{B85450}

\newtheorem{theorem}{Theorem}
\DeclareMathOperator{\fulljoin}{\tiny \textifsym{d|><|d}}
\newcommand{\system}{\emph{DeepDB}}

\vldbTitle{DeepDB: Learn from Data, not from Queries!}
\vldbAuthors{B. Hilprecht et al.}
\vldbDOI{https://doi.org/10.14778/xxxxxxx.xxxxxxx}
\vldbVolume{12}
\vldbNumber{xxx}
\vldbYear{2020}

\begin{document}

\title{DeepDB: Learn from Data, not from Queries!}

\numberofauthors{8}

\author{
\alignauthor
Benjamin Hilprecht\\
\affaddr{TU Darmstadt, Germany}
\alignauthor
Andreas Schmidt\\
\affaddr{KIT, Germany}
\alignauthor
Moritz Kulessa\\
\affaddr{TU Darmstadt, Germany}
\and  
\alignauthor
Alejandro Molina\\
\affaddr{TU Darmstadt, Germany}
\alignauthor
Kristian Kersting\\
\affaddr{TU Darmstadt, Germany}
\alignauthor
Carsten Binnig\\
\affaddr{TU Darmstadt, Germany}
}

\maketitle
\vspace*{-12ex}

\begin{abstract}
The typical approach for learned DBMS components is to capture the behavior by running a representative set of queries and use the observations to train a machine learning model. This workload-driven approach, however, has two major downsides. First, collecting the training data can be very expensive, since all queries need to be executed on potentially large databases. Second, training data has to be recollected when the workload and the data changes. 

To overcome these limitations, we take a different route: we propose to learn a pure data-driven model that can be used for different tasks such as query answering or cardinality estimation.
This data-driven model 
also supports ad-hoc queries and updates of the data without the need of full retraining when the workload or data changes.
Indeed, one may now expect that this comes at a price of lower accuracy since workload-driven models can make use of more information. 
However, this is not the case. The results of our empirical evaluation demonstrate that our data-driven approach not only provides better accuracy than state-of-the-art learned components but also generalizes better to unseen queries.
\end{abstract}

\section{Introduction}

\vspace{-3.0ex}\paragraph*{Motivation} Deep Neural Networks (DNNs) have not only been shown to solve many complex problems such as image classification or machine translation, but are applied in many other domains, too.
This is also the case for DBMSs, where DNNs have successfully been used to replace existing DBMS components with learned counterparts such as learned cost models \cite{kipf2019learned,sun2019an} as well as learned query optimizers \cite{DBLP:journals/corr/abs-1904-03711}, or even learned indexes \cite{kraska2018thecase} or query scheduling and query processing schemes \cite{DBLP:conf/sigmod/MaT19,DBLP:conf/sigmod/ShengTZP19}.

The predominant approach for learned DBMS components is that they capture the behavior of a component by running a representative set of queries over a given database and use the observations to train the model.
For example, for learned cost models such as \cite{kipf2019learned,sun2019an} different query plans need to be executed to collect the training data, which captures the runtime (or cardinalities), to then learn a model that can estimate costs for new query plans.
This observation also holds for the other approaches such as learned query optimizers or the learned query processing schemes, which are also based on collected training data that requires the execution of a representative workload. 

A major obstacle of this workload-driven approach to learning is that collecting the training data is typically very expensive since many queries need to be executed to gather enough training data.
For example, approaches like \cite{kipf2019learned,sun2019an} have shown that the runtime of hundreds of thousands of query plans is needed for the model to provide a high accuracy. 
Still, the training corpora often only cover a limited set of query patterns to avoid even higher training costs.  
For example, in \cite{kipf2019learned} the training data covers only queries up to two joins (three tables) and filter predicates with a limited number of attributes. 

\begin{figure}
	\centering
	\includegraphics[width=0.75 \linewidth]{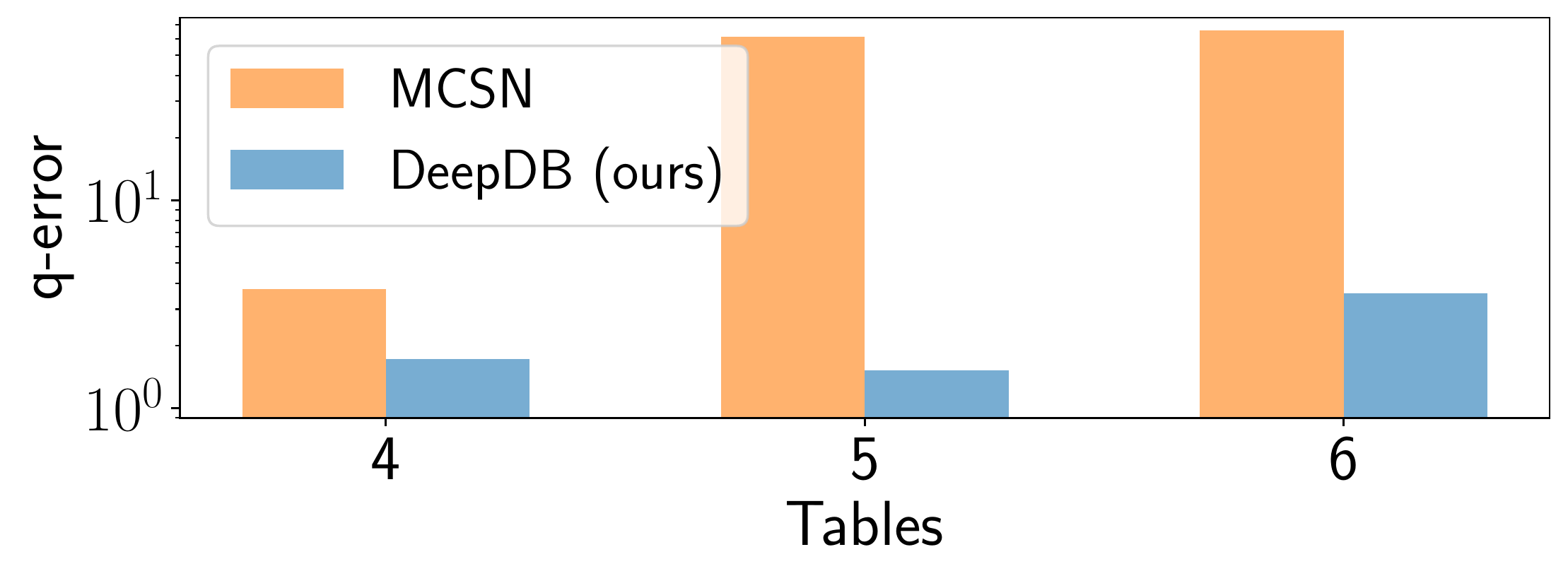}
	\vspace{-3.5ex}
  	\caption{Cardinality Estimation Errors per Join Size.}
	\vspace{-3.5ex}
	\label{fig:teaser}
\end{figure}

Moreover, the training data collection is not a one-time effort since the same procedure needs to be repeated over and over if the workload changes or if the current database is not static and the data is constantly being updated as it is typical for OLTP.
Otherwise, without collecting new training data and retraining the models for the characteristics of the changing workload or data, the accuracies of these models degrade with time. To illustrate this, 
consider Figure \ref{fig:teaser}. Here, we see (orange bars) the error of the cardinality estimation model of Kipf {\it et al.}~\cite{kipf2019learned}, called MCSN, that was trained on queries with three joined tables only.
On queries with four and more tables that the model has not seen, however, the error quickly increases.

\vspace{-1.5ex}\paragraph*{Contributions} In this paper, we take a different route. Instead of learning a model over the workload, we propose 
to learn a purely data-driven model that captures the joint probability distribution of the data and reflects important characteristics such as correlations across attributes but also the data distribution of single attributes.
Another important difference to existing approaches is that our data-density approach supports direct updates; i.e., inserts, updates, and deletes on the underlying database can be absorbed by the model without the need to retrain the model.

As a result, since our model only captures information of the data (and this is workload-independent) it can not only be used for one task but supports many different tasks ranging from query answering, over cardinality estimation to machine learning tasks such as classification or regression.
One could now think that this all comes at a price and that the accuracy of our approach must be lower since the workload-driven approaches get more information than a pure data-driven approach.
However, as we demonstrate in our experiments, this is not the case. Our approach actually outperforms many state-of-the-art workload-driven approaches.   
Furthermore, it generalizes much better.
Reconsider Figure \ref{fig:teaser}. The blue bars show the results when using our model for cardinality estimation, proving that it provides an order-of-magnitude better accuracies.

Indeed, we do not argue that data-driven models are a silver bullet to solve all possible tasks in a DBMS. 
Instead, we think that data-driven models should be combined with workload-driven models when it makes sense.
For example, a workload-driven model for a learned query optimizer might use the cardinally estimates of our model as input features.
This 
combination of data-driven 
and workload-driven models provides an interesting avenue for future work but is beyond the scope of this paper.

To summarize, the main contributions of this paper are:
\vspace{-1.5ex}
\begin{enumerate*}
\item We developed a new class of deep probabilistic models over databases, called Relational Sum Product Networks (RSPNs), that can capture important characteristics of a database.
\item To support different tasks, we devise a probabilistic query compilation approach that translates incoming database queries into probability and expectations for RSPNs (that are learned over a given database). 
\item We implemented our data-driven approach in a prototypical DBMS architecture, called \system{}, and evaluated it against state-of-the-art learned and non-learned approaches that are workload-aware, showing the benefits of our approach over these baselines.
\end{enumerate*}
\vspace{-1.5ex}

\vspace{-1.5ex}\paragraph*{Outline} 
The remainder of the paper is organized as follows.
In Section \ref{sec:overview} we first present an overview of \system{} and then discuss details of our models and the query compilation in Sections \ref{sec:relational_spns} and \ref{sec:prob_query_compilation}.
Afterwards, we explain further extensions of \system{} in Section \ref{sec:extensions} before we show an extensive evaluation comparing \system{} against state-of-the art approaches for various tasks.
Finally, we iterate over related work in Section \ref{sec:related} and conclude in Section \ref{sec:summary}.

 \section{Overview and Applications}
\label{sec:overview}

\begin{figure}
	\centering
	\includegraphics[width=0.4\textwidth]{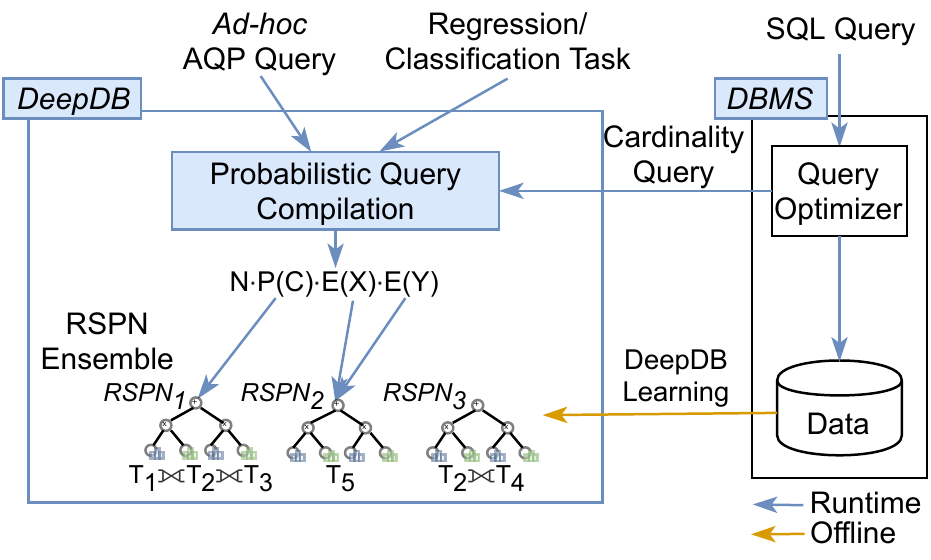}
	\vspace{-3.5ex}
	\caption{Overview of \system{}.}
	\vspace{-4.5ex}
	\label{fig:overview}
\end{figure}

\vspace{-1.5ex}
\vspace{-1.5ex}\paragraph*{Overview} As shown in Figure \ref{fig:overview}, the main idea of \system{} is to learn a 
distribution of the data.
An important aspect of \system{} is that we do not aim to replace the original data with a model (or a set of models as we discuss later).
Instead, a model created in \system{} augments a database similar to indexes to speed-up query processing and to provide additional query capabilities while we can still run standard SQL queries over the original database.

To optimally capture relevant characteristics of relational data, 
we developed a new class of models called \emph{Relational Sum Product Networks} (RSPNs).
RSPNs are based on the basic structure of Sum Product Networks (SPNs) \cite{domingos2011spn}.
In a nutshell, SPNs are deep probabilistic models that capture the joint probability distribution of a given data set.
RSPNs extend SPNs to optimize them for the use in a relational DBMS.
First, RSPNs provide additional algorithms to support a wider class of applications. Also database-specific extensions such as correct NULL-value handling etc.~are handled by RSPNs. Most importantly, what differentiates RSPNs from many other ML models, is that they support direct updates; i.e., the model does not need to be retrained but can be updated directly if new tuples are inserted into or tuples are deleted from the underlying database.

In \system{}, we create an ensemble of RSPNs that represents a given database in an offline learning procedure (similar to bulk loading an index).
Once the RSPNs are created, the models can be leveraged at runtime for different tasks.
Since RSPNs capture the joint probability distribution of the underlying database, they can support a wide variety of different applications, 
ranging from user-facing tasks (e.g., to provide fast approximate answers for SQL queries or to execute ML tasks on the model) to system-internal tasks (e.g., to provide estimates for cardinalities). 
In order to support these tasks, \system{} provides a new so called \emph{probabilistic query compilation} that translates a given task into products of expectations and probability queries on the RSPNs. 

In the following, to show that our approach is capable of supporting a range of applications, we give a brief overview of how we support these tasks in \system{}.
The main goal of this paper is to show the potentials of our data-driven learning approach to support a wide variety of different applications.
However, \system{} is not limited to the applications presented next and can be easily extended to other applications by providing a translation of queries into products of expectations and probabilities.

\vspace{-1.5ex}\paragraph*{Cardinality Estimation} The first task \system{} supports
is cardinality estimation for a query optimizer. 
Cardinality estimation is needed to provide cost estimates but also to find the correct join order during query optimization. 
Since \system{} learns a representation of the data, it can also be leveraged to provide precise cardinality estimates. 
A particular advantage of \system{} is that we do not have to create dedicated training data, i.e. pairs of queries and cardinalities.
Instead, since RSPNs capture information of the data, we can support arbitrary queries without the need to train the model for the particular workload.
Moreover, since RSPNs are easy to update they can be kept up to date at low costs similar to traditional histogram-based approaches, which is different from existing learned approaches for cardinality estimation such as \cite{kipf2019learned}.

\vspace{-1.5ex}\paragraph*{Approximate Query Processing (AQP)}
The second task we currently support in \system{} is AQP.
AQP aims to provide approximate answers to support faster query response times on large data sets. 
Currently, aggregate queries with equi-joins and typical selection predicates with group-by clauses are supported. The basic idea of how a query on a single table is executed inside \system{} is simple: for example, an aggregate query $\texttt{AVG}(X)$ with a where condition $C$ is equal to the conditional expectation $\mathbb{E}(X\mid C)$. 
These conditional expectations can be approximated with RSPNs. 
This principle can also be applied to approximate join queries. In the simplest case, a full model was learned already on the join of the corresponding tables. 
An alternative is to use multiple but smaller RSPNs that have to be combined to execute a join. All these cases are supported by our \emph{probabilistic query compilation} engine, which is explained in more detail in Section~\ref{sec:prob_query_compilation}. 

\vspace{-1.5ex}\paragraph*{Machine Learning (ML)} 
Finally, many ML tasks can also directly be conducted in \system{} based on our models without any further learning. For instance, \system{} can provide answers for regression or classification tasks for every column of the database using any set of columns as features.
 \section{Learning a Data Model}
\label{sec:relational_spns}

\begin{figure}
	\centering
	\begin{subfigure}{0.48\columnwidth}
		\centering
    	\begin{scriptsize}
        	\begin{tabular}{llll}\toprule
        		\texttt{c\_id} & \texttt{c\_age} & \texttt{c\_region} \\\midrule
        		1 & 80 & EUROPE \\
        		2 & 70 & EUROPE \\
        		3 & 60 & ASIA \\
        		4 & 20 & EUROPE \\
        		... & ... & ... \\
        		998 & 20 & ASIA \\
        		998 & 25 & EUROPE \\
        		999 & 30 & ASIA \\
        		1000 & 70 & ASIA \\\bottomrule
        	\end{tabular}
    	\end{scriptsize}
		\caption{Example Table}
		\label{fig:example_spn:table}
	\end{subfigure}
	\begin{subfigure}{0.48\columnwidth}
		\centering
    	\begin{scriptsize}
        	\begin{tabular}{ll}\toprule
        		\texttt{c\_age} & \texttt{c\_region} \\\midrule
        		\arrayrulecolor{green}
        		80 & \multicolumn{1}{|l}{EUROPE} \\
        		70 & \multicolumn{1}{|l}{EUROPE} \\
        		60 & \multicolumn{1}{|l}{ASIA} \\
        		20 & \multicolumn{1}{|l}{EUROPE} \\
        		... & \multicolumn{1}{|l}{...} \\
        		\arrayrulecolor{blue}\midrule
        		\arrayrulecolor{green}
        		... & \multicolumn{1}{|l}{...} \\
        		20 & \multicolumn{1}{|l}{ASIA} \\
        		25 & \multicolumn{1}{|l}{EUROPE} \\
        		30 & \multicolumn{1}{|l}{ASIA} \\
        		70 & \multicolumn{1}{|l}{ASIA} \\\arrayrulecolor{black}\bottomrule
        	\end{tabular}
    	\end{scriptsize}
		\caption{Learning with Row/Column Clustering}
		\label{fig:example_spn:clusters}
	\end{subfigure}
	\begin{subfigure}[t]{0.48\columnwidth}
    	\centering
    	\includegraphics[width=0.9\columnwidth]{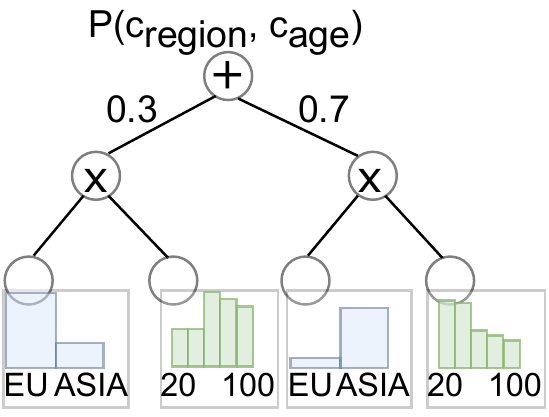}
    	\caption{Resulting SPN}
    	\label{fig:example_spn:spn}
	\end{subfigure}
	\begin{subfigure}[t]{0.48\columnwidth}
    	\centering
    	\includegraphics[width=0.9\columnwidth]{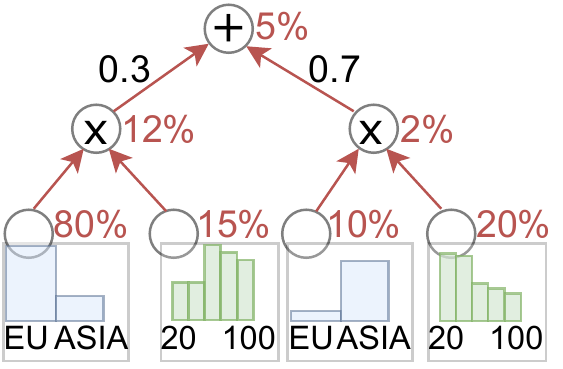}
    	\caption{Probability of European Customers younger than 30}
    	\label{fig:example_spn:prob_spn}
	\end{subfigure}
	\vspace{-1.5ex}
	\caption{Customer Table and corresponding SPN.}
	\vspace{-3.5ex}
    \label{fig:example_spn}
\end{figure}

In this section, we introduce Relational Sum Product Networks (RSPNs), which we use to learn a representation of a database and, in turn, to answer queries using our query engine explained in the next section. 
We first review 
Sum Product Networks (SPNs) and then introduce RSPNs. 
Afterwards, we describe how an ensemble of RSPNs can be created to encode a given database 
multiple tables.

\subsection{Sum Product Networks}
\label{subsec:SPN}

Sum-Product Networks (SPNs) \cite{domingos2011spn} learn the joint probability distribution $P(X_1, X_2,$ $\dots,X_n)$ of the variables $X_1,X_2,$ $\dots,$ $X_n$, which are present in the data set. They are an appealing choice because probabilities for arbitrary conditions can be computed very \emph{efficiently}. We will later make use of these probabilities for our applications like AQP and cardinality estimation.

(Tree-structured) SPNs are trees with sum and product nodes as internal nodes and leaves. Intuitively, sum nodes split the population (i.e., the rows of data set) into clusters and product nodes split independent variables of a population (i.e., the columns of a data set). Leaf nodes represent a single attribute and approximate the distribution of that attribute either using histograms for discrete domains or piecewise linear functions for continuous domains \cite{molina2017mixed}.

For instance, in Figure~\ref{fig:example_spn:spn}, an SPN was learned over the variables {\em region\/} and {\em age\/} of the corresponding {\em customer\/} table in Figure~\ref{fig:example_spn:table}. The top sum node splits the data into two groups: The left group contains $30\%$ of the population which is dominated by older European customers (corresponding to the first rows of the table) and the right group contains $70\%$ of the population with younger Asian customers (corresponding to the last rows of the table). In both groups region and age are independent and thus split by a product node each. The leaf nodes determine the probability distributions of the variables {\em region\/} and {\em age\/} for every group.

With an SPN at hand, one can compute probabilities for conditions on arbitrary columns. Intuitively, the conditions are first evaluated on every relevant leaf. Afterwards, the SPN is evaluated bottom up. 
For instance in Figure~\ref{fig:example_spn:prob_spn}, to estimate how many customers are from Europe and younger than 30, we compute the probability of European customers in the corresponding
blue {\em region\/} leaf nodes (80\% and 10\%) and the probability of a customer being younger than 30 (15\% and 20\%) in the green {\em age} leaf nodes. These probabilities are then multiplied at the product node level above, resulting in probabilities of 12\% and 2\%, respectively. Finally, at the root level (sum node), we have to consider the weights of the clusters, which leads to $12\%\cdot 0.3 + 2\%\cdot 0.7=5\%.$ Multiplied by the number of rows in the table, we get an approximation of 50 European customers who are younger than 30.

\subsection{Relational Sum-Product Networks}
Using standard SPNs directly as models for \system{} is insufficient due to the following problems: they cannot be \emph{updated} easily, leading to an obsolete data representation over time. Moreover, for our applications it is insufficient to just compute probabilities; we require \emph{extended} inference algorithms, which in particular consider \emph{database-specifics} like NULL values and functional dependencies. This led us to develop Relational SPNs (RSPNs)\footnote{Nath {\it et al.}~\cite{nath2015learning} also modified SPNs to deal with relational data. Different from RSPNs, they did neither handle updates, nor NULL-values or functional dependencies. The relational structure was exploited solely for the learning process (i.e., to avoid joining the tables before building an SPN).}.

\vspace{-1.5ex}\paragraph*{Updatability} \label{sec_update} This is the most important extensions of RS\-PNs over SPNs. If the underlying database tables are updated, the model might become inaccurate. For instance, if we insert more young European customers in the table in Figure~\ref{fig:example_spn:table}, the probability computed in Figure~\ref{fig:example_spn:prob_spn} is too low and thus the RSPN needs to be updated. 
As described before, an RSPN consists of product and sum nodes, as well as leaf nodes, which represent probability distributions for individual variables. 
The key-idea to support direct updates of an existing RSPN is to traverse the RSPN tree top-down and update the value distribution of the weights of the sum-nodes during this traversal. For instance, the weight of a sum node for a subtree of younger European customers could be increased to account for updates. Finally, the distributions in the leaf-nodes are adjusted. 
The detailed algorithm of how to directly update 
RSPNs is discussed in Section~\ref{subs:updates}.

\vspace{-1.5ex}\paragraph*{Database-specifics} First, SPNs do not provide mechanisms for handling NULL values. Hence, we developed an extension where NULL values are represented as a dedicated value for both discrete and continuous columns at the leaves during learning. Furthermore, when computing conditional probabilities and expectations, NULL values must be handled according to the three-valued logic of SQL.

Second, SPNs aim to generalize the data distribution and thus approximate the leaf distribution abstracting away speci\-fics of the data set to generalize. For instance, in the leaf nodes for the age in Figure~\ref{fig:example_spn:spn}, a piecewise linear function would be used to approximate the distribution \cite{molina2017mixed}. 
Instead, we want to represent the data as accurate as possible.
Hence, for continuous values, we store each individual value and its frequency. If the number of distinct values exceeds a given limit, we also use binning for continuous domains.

Third, functional dependencies between non-key attributes $A\rightarrow B$ are not well captured by SPNs. We could simply ignore these and learn the RSPN with both attributes $A$ and $B$ but this often leads to large SPNs since the data would be split into many small clusters (to achieve independence of $A$ and $B$). Hence, we allow users to define additional functional dependencies along with a table schema. If a functional dependency $A\rightarrow B$ is defined, we store the mapping from values of $A$ to values of $B$ in a separate dictionary of the RSPN and omit the column $B$ when learning the RSPN. At runtime, queries with filter predicates for $B$ are translated to queries with filter predicates for $A$.

\begin{figure}
	\centering
	\begin{subfigure}{0.48\columnwidth}
    	\centering
    	\includegraphics[width=0.9\columnwidth]{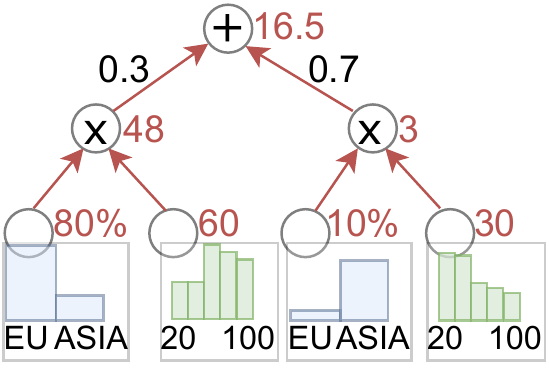}
    	\caption{$\mathbb{E}(\text{\texttt{c\_age}}\cdot 1_{\text{\texttt{c\_region='EU'}}})$}
    	\label{fig:cond_exp:ind_exp}
	\end{subfigure}
	\begin{subfigure}{0.48\columnwidth}
    	\centering
    	\includegraphics[width=0.9\columnwidth]{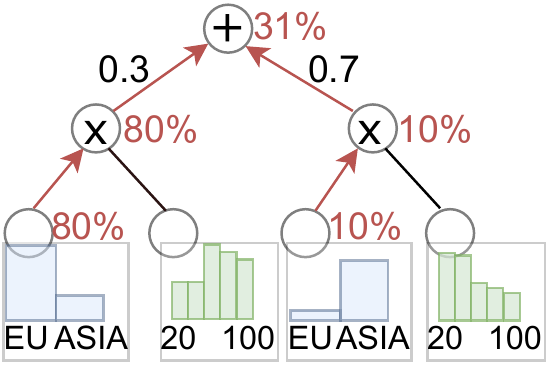}
    	\caption{$P({\text{\texttt{c\_region='EU'}}})$}
    	\label{fig:cond_exp:prob}
	\end{subfigure}
	\vspace{-1.5ex}
	\caption{Process of computing $\mathbb{E}(\text{\texttt{c\_age}}\mid{\text{\texttt{c\_region='EU'}}})$.}
	\vspace{-3.5ex}
    \label{fig:cond_exp}
\end{figure}

\vspace{-1.5ex}\paragraph*{Extended Inference Algorithms} A last important extension is that for many queries such as AVG and SUM expectations are required (e.g., to answer a SQL aggregate query which computes an average over a column). 
In order to answer these queries, RSPNs allows computing expectations over the variables on the leaves to answer those aggregates.
To additionally apply a filter predicate, we still compute probabilities on the leaves for the filter attribute and propagate both values up in the tree.
At product nodes, we multiply the expectations and probabilities coming from child nodes whereas on sum nodes the weighted average is computed. 
In Figure~\ref{fig:cond_exp} we show an example how the average age of European customers is computed. The ratio of both terms yields the correct conditional expectation. 

A related problem is that SPNs do not provide confidence intervals. We also developed corresponding extensions on SPNs 
in Section~\ref{sec:confidence_intervals}.

\subsection{Learning Ensembles of RSPNs}

An RSPN can easily be used to represent the attributes of a single table. 
However, given a more complex database with multiple tables, we have to decide which RSPNs to learn. 
Naively, one could learn a single RSPN per table.
However, then important information about dependencies between tables might be lost and lead to inaccurate approximations.
For learning an ensemble of RSPNs for a given database in \system{}, we thus take into account if tables of a schema are correlated.

\begin{figure*}
	\begin{scriptsize}
		\begin{minipage}{\textwidth}
			\begin{minipage}{.48\textwidth}
    			\begin{center}
    			    \includegraphics[height=0.13\linewidth]{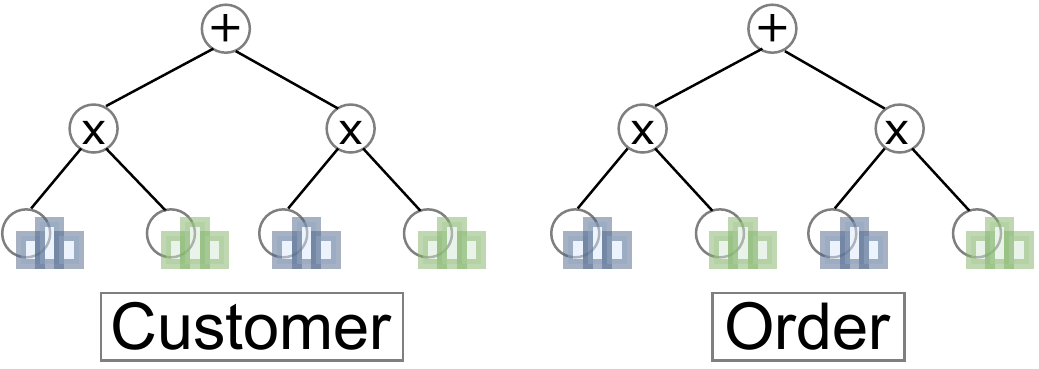}
    			\end{center}
				\begin{minipage}{0.55\textwidth}
					\texttt{Customer}\vspace{0.1em}\\
					\scriptsize
					\begin{tabular}{llll}\hline
						\texttt{c\_id} & \texttt{c\_age} & \texttt{c\_region} & {\color{blue} $\mathcal{F}_{C\leftarrow O}$} \\\hline
						1 & 20 & EUROPE & {\color{blue} 2} \\
						2 & 50 & EUROPE & {\color{blue} 0} \\
						3 & 80 & ASIA & {\color{blue} 2} \\\hline
					\end{tabular}
				\end{minipage}
				\begin{minipage}{0.45\textwidth}
					\texttt{Order}\vspace{0.1em}\\
					\scriptsize
					\begin{tabular}{llll}\hline
						\texttt{o\_id} & \texttt{c\_id} & \texttt{o\_channel} & \\\hline
						1 & 1 & ONLINE \\
						2 & 1 & STORE \\
						3 & 3 & ONLINE \\
						4 & 3 & STORE \\\hline
					\end{tabular}
				\end{minipage}
				\subcaption{Ensemble with Single Tables}
				\label{fig:prob_exp:single}
			\end{minipage}
			\hspace{0.5em}
			\begin{minipage}{.5\textwidth}
			    \centering
    			\includegraphics[height=0.13\linewidth]{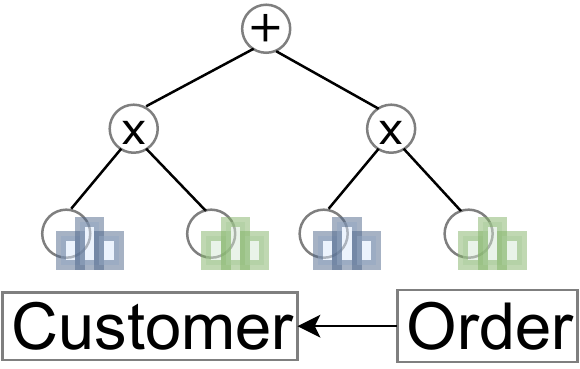}
				\begin{minipage}{1\textwidth}
				    \hspace{0.5em}\texttt{Customer}$\fulljoin$\texttt{Order}\hspace{0.1em}\\
				    \scriptsize
    				\begin{tabular}{llllllll}\hline
    					{\color{blue} $\mathcal{N}_{\mathit{C}}$} & \texttt{c\_id} & \texttt{c\_age} & \texttt{c\_region} & {\color{blue} $\mathcal{F'}_{C\leftarrow O}$} & {\color{blue} $\mathcal{N}_{\mathit{O}}$} &\texttt{o\_id} & \texttt{o\_channel}\\\hline
    					{\color{blue} 1} & 1 & 20 & EUROPE & {\color{blue} 2} & {\color{blue} 1} & 1 & ONLINE \\
    					{\color{blue} 1} & 1 & 20 & EUROPE & {\color{blue} 2} & {\color{blue} 1} & 2 & STORE \\
    					{\color{blue} 1} & 2 & 50 & EUROPE & {\color{blue} 1} & {\color{blue} 0} & NULL & NULL \\
    					{\color{blue} 1} & 3 & 80 & ASIA & {\color{blue} 2} & {\color{blue} 1} & 3 & ONLINE \\
    					{\color{blue} 1} & 3 & 80 & ASIA & {\color{blue} 2} & {\color{blue} 1} & 4 & STORE \\\hline
    				\end{tabular}
    				\subcaption{Ensemble with Full Outer Join}
    				\label{fig:prob_exp:join}
				\end{minipage}
			\end{minipage}
		\end{minipage}

		\vspace{-1.5ex}
		\caption{Two RSPN Ensembles for the same Schema. Additional (blue) columns are also learned by the RSPNs.}
		\vspace{-3.5ex}
		\label{fig:prob_exp}
	\end{scriptsize}
\end{figure*}

In the following, we describe our procedure that constructs a so called \emph{base ensemble} for a given database scheme.
In this procedure, for every {\em foreign key\/}$\rightarrow${\em primary key\/} relationship we learn an RSPN over the corresponding full outer join of two tables if there is a correlation between attributes of the different tables. Otherwise, RSPNs for the single tables will be learned. For instance, if the schema consists of a \texttt{Customer} and an \texttt{Order} table as shown in Figure \ref{fig:prob_exp}, we could either learn two independent RSPNs (one for each table) or a joint RSPN (over the full outer join).
In order to test independence of two tables and thus to decide if one or two RSPNs are more appropriate, we check for every pair of attributes from these tables if they can be considered independent or not. In order to enable an efficient computation, this test can be done on a small random sample.
As a correlation measure that does not make major distributional assumptions, we compute RDC values \cite{lopez2013randomized} between two attributes, which are also used in the MSPN learning algorithm \cite{molina2017mixed}.
If the maximum pairwise RDC value between all attributes of two tables exceeds a threshold (where we use the standard thresholds of SPNs), we assume that two tables are correlated and learn an RSPN over the join. 
Otherwise, single RSPNs are learned.  

In the base ensemble only correlations between two tables are captured. While in our experiments, we see that this already leads to highly accurate answers, there might also be correlations not only between directly neighboring tables. Learning these helps to further improve the accuracy of queries that span more than two tables.
For instance, if there was an additional \texttt{Product} table that can be joined with the \texttt{Orders} table and the product prize is correlated with the customers region, this would not be taken into account in the \emph{base ensemble}.  
In Section \ref{sec:ensemble_optimization}, we extend our basic procedure for ensemble creation to take dependencies among multiple tables into account.

 \section{Probabilistic Query Compilation}
\label{sec:prob_query_compilation}

The main challenge of probabilistic query compilation is to translate an incoming query (e.g., for AQP) into an inference procedure against an ensemble of RSPNs. To this end, 
recall that an ensemble for a given database either consists of RSPNs for single tables or spanning two (or more) tables.

In the following, we first describe how the translation procedure for a COUNT query works (which can be used either for AQP or for cardinality estimation) and then extend it to more complex queries (e.g., AVG and SUM).
We then show how machine learning tasks can be supported with the help of RSPNs.

\subsection{COUNT Queries}
\label{sub:count_queries}

In this section, we explain how we can translate simple COUNT queries with and without filter predicates over single tables as well as COUNT queries that join multiple tables using inner joins (equi-joins).
For filter predicates we support conjunctions of predicates of the form $a\,op\,c$ where $a$ is an attribute, $c$ a constant, and $op$ one of the comparison operators $>, < , \geq, \leq, \neq$ or an IN-comparison (e.g., age IN (20, 30, 40)). String or arithmetic expressions, as well as user-defined functions are currently not supported.
Disjunctions could be realized using the inclusion-exclusion principle.

These types of queries can be used already for cardinality estimation but also cover some cases of aggregate queries for AQP.
We later show the extensions to support a broader set of queries for AQP including other aggregates (AVG and SUM) as well as group-by statements.
For answering the simple COUNT queries, we distinguish three cases of how queries can be mapped to RSPNs: (1) an RSPN exists that exactly matches the tables of the query, (2) the RSPN is larger and covers more tables, and (3) we need to combine multiple RSPNs since there is no single RSPN that contains all tables of the query.

\paragraph*{Case 1: Exact matching RSPN available} The simplest case is a single table COUNT query with (or without) a filter predicate. If an RSPN is available for this table and $N$ denotes the  number of rows in the table, the result is simply $N \cdot P(C)$. For instance, the query
\begin{lstlisting}
$Q_1$: SELECT COUNT(*) 
      FROM CUSTOMER C
     WHERE c_region='EUROPE';
\end{lstlisting}
can be answered with the \texttt{CUSTOMER} RSPN in Figure \ref{fig:prob_exp:single}. The result is
$|\texttt{C}|\cdot\mathbb{E}(\mathbf{1}_{\texttt{c\_region='EUROPE'}})=3\cdot \frac{2}{3}=2$. Note that $1_C$ denotes the random variable being one if the condition $C$ is fulfilled and thus $\mathbb{E}(\mathbf{1}_{C})=P(C)$.

A natural extension for COUNT queries over joins could be to learn an RSPN for the underlying join and use the formula $|J|\cdot P(C)$ where the size of the joined tables without applying a filter predicate is $|J|$. 
For instance, the query
\begin{lstlisting}
$Q_2$: SELECT COUNT(*) 
      FROM CUSTOMER C 
   NATURAL JOIN ORDER O
     WHERE c_region='EUROPE' 
       AND o_channel='ONLINE';
\end{lstlisting}
could be represented as $|\texttt{C}\bowtie\texttt{O}|\cdot P(\texttt{o\_channel='ONLINE'} \cap \texttt{c\_region='EUROPE'})$
which is $4\cdot \frac{1}{4}=1$.

However, joint RSPNs over multiple tables are learned over the full outer join.
By using full outer joins we preserve all tuples of the original tables and not only those that have one or more join partner in the corresponding table(s). This way we are able for example to answer also single table queries from a joint RSPN, as we will see in Case 2.
The additional NULL tuples that result from a full outer join must be taken into account when answering an inner join query. For instance, the second customer in Figure \ref{fig:prob_exp:join} does not have any orders and thus should not be counted for query $Q_2.$
To make it explicit which tuples have no join partner and thus would not be in the result of an inner join, we add an additional column $\mathcal{N}_T$ for every table such as in the ensemble in Figure \ref{fig:prob_exp:join}. 
This column is also learned by the RSPN and can be used as an additional filter column to eliminate tuples that do not have a join partner for the join query given. Hence, the complete translation of query $Q_2$ for the RSPN learned over the full outer join in Figure~\ref{fig:prob_exp:join} is $|\texttt{C}\fulljoin\texttt{O}|\cdot P(\texttt{o\_channel='ONLINE'} \cap \texttt{c\_region='EUROPE'} \cap \mathcal{N}_O=1 \cap \mathcal{N}_C=1)=5\cdot \frac{1}{5}=1$.

\paragraph*{Case 2: Larger RSPN available}  The second case is that we have to use an RSPN that was created on a set of joined tables, however, the query only needs a subset of those tables.
For example, let us assume that the query $Q_1$ asking for European customers is approximated using an RSPN learned over a full outer join of customers and orders such as the one in Figure \ref{fig:prob_exp:join}. 
The problem here is that customers with multiple orders would appear several times in the join and thus be counted multiple times. For instance, the ratio of European customers in the full outer join is $3/5$ though two out of three customers in the 
data set 
are European. 

To address this issue, for each {\em foreign key\/}$\rightarrow${\em primary key\/} relationship $S \leftarrow P$ between tables $P$ and $S$ we add a column $\mathcal{F}_{S \leftarrow P}$ to table $S$ denoting how many corresponding join partners a tuple has. We call these \emph{tuple factors} and later use them as correction factor. 
For instance, in the customer table in Figure~\ref{fig:prob_exp:single} for the first customer the tuple factor is two since there are two tuples in the order table for this customer. 
It is important to note that tuple factors have to be computed only once per pair of tables that can be joined via a foreign key.
In \system{}, we do this when the RSPNs for a given database are created and our update procedure changes those values as well. 
Tuple factors are included as additional column and learned by the RSPNs just as usual columns.
When used in a join, we denote them as $\mathcal{F'}_{S \leftarrow P}$. Since we are working with outer joins, the value of $\mathcal{F'}$ is at least 1.

We can now express the query that asks for the count of customers from Europe as 
\begin{equation*}
|\texttt{C}\fulljoin\texttt{O}|\cdot\mathbb{E}\left(\frac{1}{\mathcal{F'}_{C\leftarrow O}}\cdot\mathbf{1}_{\texttt{c\_region='EUROPE'}}\cdot\mathcal{N}_C\right)
\end{equation*}
which results in $5\cdot\frac{1/2+1/2+1}{5}=2$. 
First, this query both includes the first customer (who has no orders) because the RSPN was learned on the full outer join.
Second, the query also takes into account that the second and third customer have two orders each by normalizing them with their tuple factor  $\mathcal{F'}_{C \leftarrow O}$.

In general, 
we can define the procedure to compile a query requiring only a part of an RSPN as follows:
\begin{theorem}
\label{theo:larger_rspn}
	Let $Q$ be a \texttt{COUNT} query with a filter predicate $C$ which only queries a subset of the tables of a full outer join $J.$ Let $\mathcal{F'}(Q,J)$ denote the product of all tuple factors that cause result tuples of $Q$ to appear multiple times in $J.$
	The result of the query is equal to: 
	\begin{equation*}
		|J|\cdot\mathbb{E}\left(\frac{1}{\mathcal{F'}(Q,J)}\cdot 1_{C} \cdot \displaystyle \prod_{T\in Q}\mathcal{N}_{T}\right)
	\end{equation*}
\end{theorem}
For an easier notation, we write the required factors of query $Q$ as $\mathbf{F}(Q)$. The expectation $\mathbb{E}(\mathbf{F}(Q))$ of theorem \ref{theo:larger_rspn} can be computed with an RSPN because all columns are learned.

\paragraph*{Case 3: Combination of multiple RSPNs} As the last case, we handle a \texttt{COUNT} query that needs to span over multiple RSPNs. We first handle the case of two RSPNs and extend the procedure to $n$ RSPNs later.
In this case, the query can be split into two subqueries $Q_L$ and $Q_R$, one for each RSPN. There can also be an overlap between $Q_L$ and $Q_R$ which we denote as $Q_O$ (i.e., a join over the shared common tables). The idea is first to estimate the result of $Q_L$ using the first RSPN. We then multiply this result by the ratio of tuples in $Q_R$ vs. tuples in the overlap $Q_O$. Intuitively, this expresses how much the missing tables not in $Q_L$ increase the COUNT value of the query result.

For instance, there is a separate RSPN available for the \texttt{Customer} and the \texttt{Order} table in Figure~\ref{fig:prob_exp:single}. The query $Q_2$, as shown before, would be split into two queries $Q_{L}$ and $Q_{R}$, one against the RSPN built over the \texttt{Customer} table and the other one over the RSPN for the \texttt{Order} table. $Q_O$ is empty in this case. The query result of $Q_2$ can thus be expressed using all these sub-queries as: 
\begin{equation*}
|\texttt{C}|
\cdot
\underbrace{\mathbb{E}(\mathbf{1}_{\texttt{c\_region='EUROPE'}}\cdot\mathcal{F}_{C\leftarrow O})}_{Q_L}
\cdot
\underbrace{\mathbb{E}(\mathbf{1}_{\texttt{o\_channel='ONLINE'}})}_{Q_R}
\end{equation*}
which results in $3\cdot\frac{2+0}{3}\cdot\frac{2}{4}=1.$ The intuition of this query is that the left-hand side that uses $Q_L$ computes the orders of European customers while the right-hand side computes the fraction of orders that are ordered online out of all orders.

We now handle the more general case that the overlap is not empty and that there is a foreign key relationship $S\leftarrow T$ between a table $S$ in $Q_O$ (and $Q_L$) and a table $T$ in $Q_R$ (but not in $Q_L$). In this case, we exploit the tuple factor $\mathcal{F}_{S\leftarrow T}$ in the left RSPN. We now do not just estimate the result of $Q_L$ but of $Q_L$ joined with the table $T.$ Of course this increases the overlap which we now denote as $Q_O'.$ As a general formula for this case, we obtain Theorem \ref{theo:join}:
\begin{theorem}
	\label{theo:join}
	Let the filter predicates and tuple factors of $Q_L \setminus Q_O$ and $Q_R \setminus Q_O$ be conditionally independent given the filter predicates of $Q_O$. Let $S\leftarrow T$ be the foreign key relationship between a table $S$ in $Q_{L}$ and a table $T$ in $Q_{R}$ that we want to join. The result of $Q$ is equal to 
	\begin{equation*}
		|J_L|\cdot\mathbb{E}\left(\mathbf{F}(Q_L) \cdot \mathcal{F}_{S\leftarrow T}\right)
		\cdot
		\frac{\mathbb{E}\left(\mathbf{F}(Q_R)\right)}{\mathbb{E}\left(\mathbf{F}(Q_O')\right)}.
	\end{equation*}
\end{theorem}

Independence across RSPNs is often given since our ensemble creation procedure preferably learns RSPNs over correlated tables as discussed in Section \ref{sec:relational_spns}. 

Alternatively, we can start the execution with $Q_R$.
In our example query $Q_2$ where $Q_R$ is the query over the orders table, we can remove the corresponding tuple factor $\mathcal{F}_{C\leftarrow O}$ from the left expectation. 
However, we then need to normalize $Q_L$ by the tuple factors to correctly compute the fraction of customers who come from Europe.
To that end, the query $Q_2$ can alternatively be computed using:
\begin{equation*}
|\texttt{O}|\cdot\mathbb{E}(\mathbf{1}_{\texttt{o\_channel='ONLINE'}})
\cdot 
\frac{\mathbb{E}\left(\mathbf{1}_{\texttt{c\_region='EUROPE'}}
\cdot
\mathcal{F}_{C\leftarrow O}\mid \mathcal{F}_{C\leftarrow O}\right)}{\mathbb{E}\left(\ \mathcal{F}_{C\leftarrow O}\mid \mathcal{F}_{C\leftarrow O}>0\right)}
\end{equation*}

\vspace{-1.5ex}\paragraph*{Execution Strategy} If multiple RSPNs are required to answer a query, we have several possible execution strategies. Our goal should be to handle as many correlations between filter predicates as possible because predicates across RSPNs are considered independent. For instance, assume we have both the \texttt{Customer}, \texttt{Order} and \texttt{Customer-Order} RSPNs of Figure~\ref{fig:prob_exp} in our ensemble, and a join of customers and orders would have filter predicates on  $\texttt{c\_region}, $ $\texttt{c\_age}$ and $\texttt{o\_channel}.$ In this case, we would prefer the \texttt{Customer-Order} RSPN because it can handle all pairwise correlations between filter columns (\texttt{c\_region}-\texttt{c\_age}, \texttt{c\_region}-\texttt{o\_channel}, \texttt{c\_age}-\texttt{c\_channel}).
Hence, at runtime we greedily use the RSPN that currently handles the filter predicates with the highest sum of pairwise RDC values. We also experimented with strategies enumerating several probabilistic query compilations and using the median of their predictions. However, this was not superior to our RDC-based strategy. Moreover, the RDC values have already been computed to decide which RSPNs to learn. Hence, at runtime this strategy is very compute-efficient.

The final aspect is how to handle joins spanning over more than two RSPNs.
To support this, we can apply Theorem~\ref{theo:join} several times. 

\subsection{Other AQP Queries}

So far, we only looked into COUNT queries without group-by statements.
In the following, we first discuss how we extend our query compilation to also support AVG and SUM queries before we finally explain group-by statements as well as outer joins.

\vspace{-1.5ex}\paragraph*{AVG Queries} We again start with the case that we have an RSPN that exactly matches the tables of a query and later discuss the other cases. For this case, queries with \texttt{AVG} aggregates can be expressed as conditional expectations. For instance, the query
\begin{lstlisting}
$Q_3$: SELECT AVG(c_age) 
      FROM CUSTOMER C
     WHERE c_region='EUROPE';
\end{lstlisting}
can be formulated as $|\texttt{C|}\cdot\mathbb{E}(\texttt{c\_age}\mid \texttt{c\_region='EUROPE'})$ with the ensemble in Figure~\ref{fig:prob_exp:single}.

However, for the case that an RSPNs spans more tables than required, we cannot directly use this conditional expectation because otherwise customers with several orders would be weighted higher. Again, normalization by the tuple factors is required. 
For instance, if the RSPN spans customers and orders as in Figure~\ref{fig:prob_exp:join} for query $Q_3$ we use
\begin{equation*}
    \frac{\mathbb{E}\left(\frac{\texttt{c\_age}}{\mathcal{F'}_{C\leftarrow O}}\mid \texttt{c\_region='EUROPE'}\right)}{\mathbb{E}\left(\frac{1}{\mathcal{F'}_{C\leftarrow O}}\mid \texttt{c\_region='EUROPE'}\right)}=\frac{20/2+20/2+50}{1/2+1/2+1}=35.
\end{equation*}

In general, if an average query for the attribute $A$ should be computed for a join query $Q$ with filter predicates $C$ on an RSPN on a full outer join $J$, we use the following expectation to answer the average query:
\begin{equation*}
    \mathbb{E}\left(\frac{A}{\mathcal{F'}(Q,J)}\mid C\right)/\mathbb{E}\left(\frac{1}{\mathcal{F'}(Q,J)}\mid C\right).
\end{equation*}

The last case is where the query needs more than one RSPN to answer the query.
In this case, we only use one RSPN that contains $A$ and ignore some of the filter predicates that are not in the RSPN.
As long as $A$ is independent of these attributes, the result is correct. Otherwise, this is just an approximation. 
For selecting which RSPN should be used, we again prefer RSPNs handling stronger correlations between $A$ and $P$ quantified by the RDC values.
The RCDs can also be used to detect cases where the approximation would ignore strong correlations with the missing attributes in $P$.

\vspace{-1.5ex}\paragraph*{SUM Queries} For handling \texttt{SUM} queries we run two queries: one for the \texttt{COUNT} and \texttt{AVG} queries. Multiplying them yields the correct result for the \texttt{SUM} query.

\vspace{-1.5ex}\paragraph*{Group-by Queries} Finally, a \texttt{group by} query can be handled also by several individual queries with additional filter predicates for every group. This means that for $n$ groups we have to compute $n$ times more expectations than for the corresponding query without grouping. In our experimental evaluation, we show that this does not cause performance issues in practice if we compute the query on the model.

\vspace{-1.5ex}\paragraph*{Outer Joins} Query compilation can be easily extended to support outer joins (left/right/full). The idea is that we only filter out tuples that have no join partner for all inner joins (case 1 and 2 in Section \ref{sub:count_queries}) but not for outer joins (depending on the semantics of the outer join). Moreover, in case 3, the tuple factors $\mathcal{F}$ with value zero have to be handled as value one to support the semantics of the corresponding outer join.

\subsection{Machine Learning (ML) Tasks}

Many ML tasks can directly also be expressed using RSPNs.
For example, regression tasks can directly be translated into conditional expectations. For classification we can use most probable explanation (MPE) algorithms \cite{molina2017mixed}. RSPNs are optimized to accurately represent the data which is beneficial for AQP and cardinality estimation. However, they still generalize since the dependency structure of the data is identified and thus the regression and classification performances are competitive as we show in our experiments.

 \section{DeepDB Extensions}
\label{sec:extensions}

We now describe important extensions of our basic framework presented before. We first explain how confidence intervals are provided, which is especially important for AQP.
We then discuss how RSPNs can be updated if the database is changed. 
Finally, we present how we can optimize the basic ensemble of RSPNs by additional RSPNs that can span more than two tables. 

\subsection{Support for Confidence Intervals}
\label{sec:confidence_intervals}

Especially for AQP confidence intervals are important. However, SPNs do not provide those. After the probabilistic query compilation the query is expressed as a product of expectations. We first describe how to estimate the uncertainty for each of those factors and eventually how a confidence interval for the final estimate can be derived.

First, we split up expectations as a product of probabilities and conditional expectations. For instance, the expectation $\mathbb{E}(X \cdot 1_C)$ would be turned into $\mathbb{E}(X\mid C)\cdot P(C)$. This allows us to treat all probabilities for filter predicates $C$ as a single binomial variable with probability $p=\displaystyle \prod P(C_i)$ and the amount of training data of the RSPN as $n_{\mathit{samples}}$. Hence, the variance is $\sqrt{n_{\mathit{samples}} p (1-p)}$. For the conditional expectations, we use the Koenig-Huygens formula $\mathbb{V}(X\mid C)=\mathbb{E}(X^2\mid C)-\mathbb{E}(X\mid C)^2$. Note that also squared factors can be computed with RSPNs since the square can be pushed down to the leaf nodes. We now have a variance for each factor in the result.

For the combination we need two simplifying assumptions: (i) the estimates for the expectations and probabilities are independent, and (ii) the resulting estimate is normally distributed. In our experimental evaluation, we show that despite these assumptions our confidence intervals match those of typical sample-based approaches. 

We can now approximate the variance of the product using the independence assumption by recursively applying the standard equation for the product of independent random variables: $\mathbb{V}(XY)=\mathbb{V}(X)\mathbb{V}(Y)+\mathbb{V}(X)\mathbb{E}(Y)^2+\mathbb{V}(Y)\mathbb{E}(X)^2$.
Since we know the variance of the entire probabilistic query compilation and we assume that this estimate is normally distributed we can provide confidence intervals.

\subsection{Support for Updates}
\label{subs:updates} 

The intuition of our update algorithm is to regard RSPNs as indexes. Similar to these, insertions and deletions only affect subtrees and can be performed recursively. Hence, the updated tuples recursively traverse the tree and passed weights of sum nodes and the leaf distributions are adapted. Our approach supports {\em insert\/} and {\em delete\/} operations, where an {\em update\/}-operation is mapped to a pair of {\em delete\/} and {\em insert\/} operations.

\begin{algorithm}
\begin{scriptsize}
\caption{Incremental Update}\label{alg:inc_learn}
\begin{algorithmic}[1]
\Procedure{update\_tuple}{$node, tuple$}
\If{leaf-node} \label{leafNode}
    \State update\_leaf\_distribution($node, tuple$) 
\ElsIf{sum-node}  \label{sumNode}
   \State $nearest\_child \gets get\_nearest\_cluster(node, tuple)$ \label{nearest_cluster}
   \State adapt\_weights($node, nearest\_child$) \label{adjust_weights}
   \State update\_tuple($ nearest\_child, tuple$) \label{recursive_call_on_child}
\ElsIf{product-node} \label{productNode}
   \For {$child$ in $child\_nodes$}                \label{iterate_childs_start}
   	\State $tuple\_proj\leftarrow project\_to\_child\_scope(tuple)$
     \State update\_tuple($child, tuple\_proj$)  \label{iterate_childs_end}
   \EndFor
\EndIf
\EndProcedure
\end{algorithmic}
\end{scriptsize}
\end{algorithm}

The update algorithm is summarized in Algorithm~\ref{alg:inc_learn}. Since it is recursive, we have to handle sum, product and leaf nodes. At sum nodes (line~\ref{sumNode}) we have to identify to which child node the inserted (deleted) tuple belongs to determine which weight has to be increased (decreased). Since children of sum nodes represent row clusters found by {\em KMeans\/} during learning \cite{molina2017mixed}, we can compute the closest cluster center (line~\ref{nearest_cluster}), increase (decrease) its weight (line~\ref{adjust_weights}) and propagate the tuple to this subtree (line~\ref{recursive_call_on_child}). In contrast, product nodes (line~\ref{productNode}) split the set of columns. Hence, we do not propagate the tuple to one of the children but split it and propagate each tuple fragment to the corresponding child node (lines~\ref{iterate_childs_start}-\ref{iterate_childs_end}). Arriving at a leaf node, only a single column of the tuple is remaining. We now update the leaf distribution according to the column value (line~\ref{leafNode}).

This approach does not change the structure of the RSPN, but only adapts the weights and the histogram values. If there are new dependencies as a result of inserts they are not represented in the RSPN. As we show in Section~\ref{para:update-results} on a real-word data set, this typically does not happen, even for high incremental learning rates of 40\%. Nevertheless, in case of new dependencies the RSPNs have to be rebuilt. This is solved by checking the database cyclically for changed dependencies by calculating the pairwise RDC values as explained in Section~\ref{sec:ensemble_optimization} on column splits of product nodes. If changes are detected in the dependencies, the affected RSPNs are regenerated. As for traditional indexes, this can be done in the background. 
\subsection{Ensemble Optimization}
\label{sec:ensemble_optimization}

As mentioned before, we create an ensemble of RSPNs for a given database.
The base ensemble contains either RSPNs for single tables or they span over two tables connected by a foreign key relationship if they are correlated.
Correlations occurring over more than two tables are ignored so far since they lead to larger models and higher training times.
In the following, we thus discuss an extension of our ensemble creation procedure that allows a user to specify a training budget (in terms of time or space) and \system{} selects the additional larger RSPNs that should be created. 
We formulate the problem of which additional RSPNs to learn as constrained optimization problem. 

\begin{figure}
	\centering
	\begin{subfigure}[b]{0.9\linewidth}
		\centering
		\includegraphics[width=0.7\linewidth]{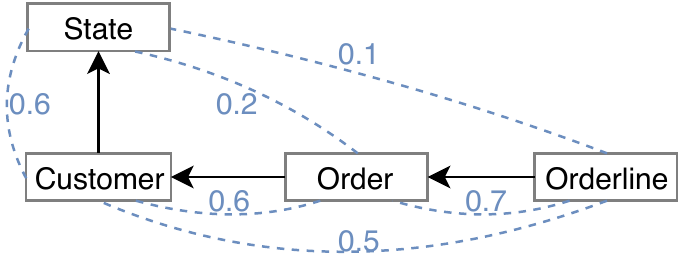}
		\caption{Schema with pairwise Dependencies (RDC)}
		\label{fig:example_ensemble:rdc_values}
	\end{subfigure}
	\begin{subfigure}[b]{0.95\linewidth}
		\centering
		\includegraphics[width=1\linewidth]{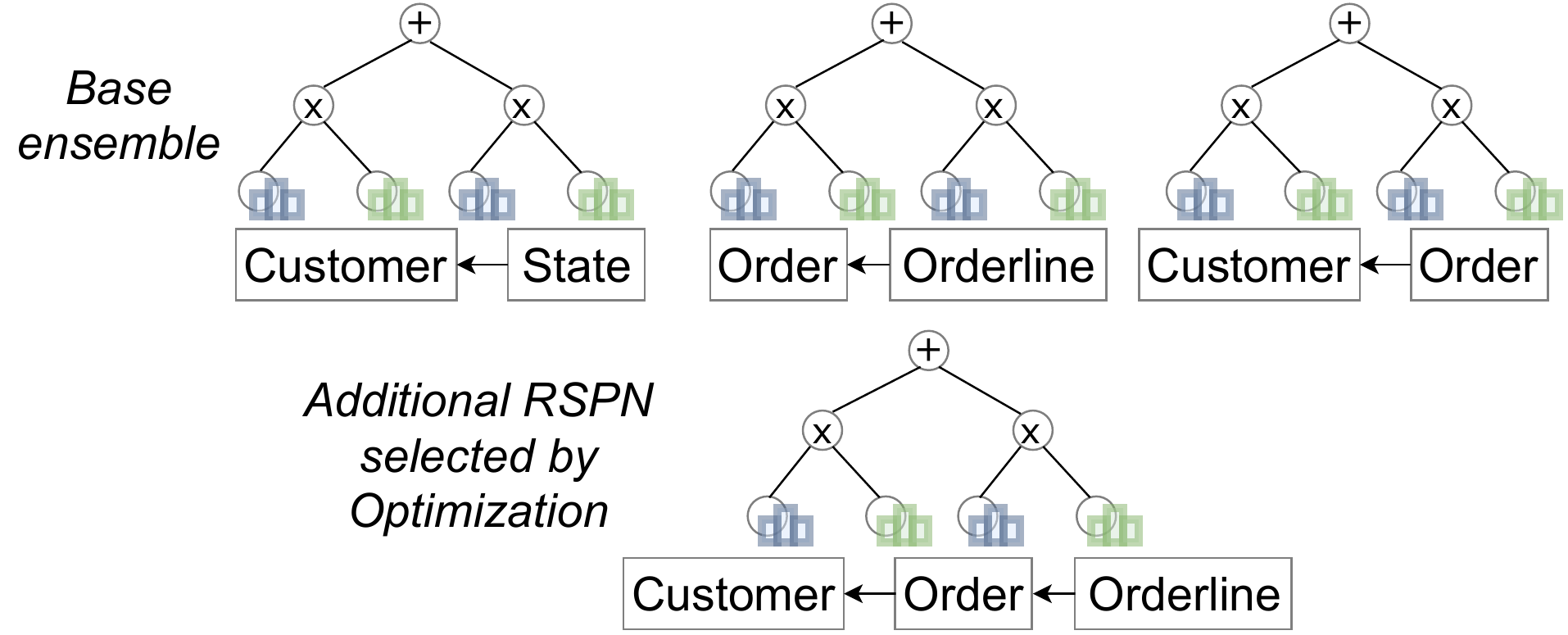}
		\caption{Base RSPN Ensemble with RDC threshold 0.3 and additional RSPN selected by Optimization}
	\end{subfigure}
	\vspace{-1.5ex}
	\caption{RSPN ensemble}
	\vspace{-3.5ex}
	\label{fig:example_ensemble}
\end{figure}

To quantify the correlations between tables, as mentioned already before, we compute the pairwise RDC values for every pair of attributes in the schema. For every pair of tables, we define the maximum RDC value between two columns $\max_{c\in T_i, c'\in T_j} \mathit{rdc}(c,c')$ as the dependency value. 
The dependency value indicates which tables should appear in the same RSPN and which not.
An example is given in Figure~\ref{fig:example_ensemble:rdc_values}. Here, the \texttt{Customer}, \texttt{Order} and \texttt{Orderline} tables have high pairwise correlations while the \texttt{State} table is only highly correlated with the \texttt{Customer} table.

For every RSPN the goal is to achieve a high mean of these pairwise maximal RDC values. This ensures that only tables with high pairwise correlation are merged in an RSPN. 
For instance, the mean RDC value for the RSPN learned over the full outer join of the tables \texttt{Customer}, \texttt{Order} and \texttt{Orderline} would be $(0.6+0.7+0.5)/3=0.6$. This RSPN is more valuable than an RSPN learned over the \texttt{State}, \texttt{Customer} and \texttt{Order} tables with a mean RDC value of  $(0.6+0.6+0.2)/3=0.46$. The overall objective function for our optimization procedure to create an ensemble is thus to maximize the sum of all mean RDC values of the RSPNs in the ensemble.

The limiting factor (i.e., the constraint) for the additional RSPN ensemble selection should be the budget (i.e., extra time compared to the base ensemble) we allow for the learning of additional RSPNs. 
For the optimization procedure, we define the maximum learning costs as a factor $B$ relative to the learning costs of the base ensemble $C_{\mathit{Base}}$. Hence, a budget factor $B=0$ means that only the base ensemble would be created. For higher budget factors $B>0$, additional RSPNs over more tables are learned in addition.
If we assume that an RSPN $r$ among the set of all possible unique RSPNs $R$ has a cost $C(r)$, then we could formulate the optimization problem as follows:

\vspace{-2ex}
\begin{equation*}
\begin{aligned}
& \underset{\mathcal{E}}{\text{minimize}}
& & \sum_{r\in \mathcal{E}} \overline{\{\max_{c\in T_i, c'\in T_j} \mathit{rdc}(c,c') \mid T_i, T_j \in r\}}\\
& \text{subject to}
& & \sum_{r\in \mathcal{E}} C(r)\le B\cdot C_{\mathit{Base}}
\end{aligned}
\end{equation*}

However, estimating the real cost $C(r)$  (i.e., time) to build an RSPN $r$ is hard and thus we can not directly solve the optimization procedure.
Instead, we estimate the relative cost to select the RSPN $r$ that has the highest mean RDC value and the lowest relative creation cost.
To model the relative creation cost, we assume that the costs grow quadratic with the number of columns $cols(r)$ since the RDC values are created pairwise and linear in the number of rows $rows(r)$. Consequently, we pick the RSPN $r$ with highest mean RDC and lowest cost which is $cols(r)^2 \cdot rows(r)$ as long as the maximum training time is not exceeded.

 \section{Experimental Evaluation}

In this Section, we show that \system{} outperforms state-of-the-art systems for both cardinality estimation and AQP, where we not only demonstrate the performance of \system{} for both tasks but also show the capabilities of updating RSPNs. 
Moreover, we also study the performance of \system{} for different ML tasks.

The RSPNs we used in all experiment were implemented in Python as extensions of SPFlow \cite{molina2019spflow}. As hyperparameters, we used an RDC threshold of $0.3$ and a minimum instance slice of $1\%$ of the input data, which determines the granularity of clustering. Moreover, we used a budget factor of 0.5, i.e. the training of the larger RSPNs takes approximately 50\% more training time than the base ensemble. 
We determined the hyperparameters using a grid-search, which gave us the best results across different data sets.

\subsection{Exp. 1: Cardinality Estimation}

First, we compare the prediction quality of \system{} which is purely data-driven with state-of-the-art learned cardinality estimation techniques that take the workload into account.

\vspace{-1.5ex}\paragraph*{Baselines} In addition to the learned baselines, we also compare against non-learned baselines. First we trained a Multi-Set Convolutional Network (MCSN) \cite{kipf2019learned} as a learned baseline. MCSNs are specialized deep neural networks using the join paths, tables and filter predicates as inputs. Also, we use Index-Based Join Sampling \cite{leis2017ibjs} as a non-learned baseline. This algorithm exploits secondary indexes to estimate the full join size using sampling.  Furthermore, the standard cardinality estimation of Postgres 11.5 was employed as another non-learned baseline. Additionally, we implemented random sampling.

\vspace{-1.5ex}\paragraph*{Workload}  As in~\cite{kipf2019learned, leis2015how}, the JOB-light benchmark is used. The benchmark uses the real-world IMDb database and  defines 70 queries. Furthermore, we additionally defined a synthetic query set of 200 queries were joins from three to six tables and one to five filter predicates appear uniformly on the IMDb data set. We use this query set to compare the generalization capabilities of the learned approaches.

\vspace{-1.5ex}\paragraph*{Training Time} In contrast to other learned approaches for cardinality estimation \cite{kipf2019learned, sun2019an}, no dedicated training data is required for \system{}. Instead, we just learn a representation of the data. The training of the base ensemble takes 48 minutes. 
The creation time includes the data preparation time to compute the tuple factors as introduced in Section~\ref{sub:count_queries}.

In contrast, for the MCSN \cite{kipf2019learned} approach, 100k queries need to be executed to collect cardinalities resulting in 34 hours of training data preparation time (when using Postgres). Moreover, the training of the neural network takes about 15 minutes on a Nvidia V100 GPU. 

We see that our training time is much lower since we do not need to collect any training data for the workload.
Another advantage is that we do not have to re-run the queries once the database is modified.
Instead, we provide an efficient algorithm to update RSPNs in \system{} as discussed in Section \ref{sec_update}.

\begin{table}
    \scriptsize
	\centering
	\begin{tabular}{lllll}\toprule
		& median & 90th & 95th & max \\\midrule
		DeepDB (ours)& \underline{1.27} & \underline{2.50} & \underline{3.16} & \underline{39.66} \\
MCSN & 3.22 & 65 & 143 & 717 \\
 		Postgres & 6.84 & 162 & 817 & 3477 \\
		IBJS & 1.67 & 72 & 333 & 6949 \\
		Random Sampling & 5.05 & 73 & 10371 & 49187 \\\bottomrule
	\end{tabular}
	\vspace{-1.5ex}
	\caption{Estimation Errors for the JOB-light Benchmark}
	\vspace{-4.5ex}
	\label{fig:q_errors}
\end{table}

\vspace{-1.5ex}\paragraph*{Estimation Quality} The prediction quality of cardinality estimators is usually evaluated using the q-error which is the factor by which an estimate differs from the real execution join size. 
For example, if the real result size of a join is 100, the estimates of 10 or 1000 tuples both have a q-error of 10. 
Using the ratio instead of an absolute or quadratic error captures the intuition that for making optimization decisions only relative differences matter.

In Table \ref{fig:q_errors} we depicted the median, 90-th and 95-th percentile and max q-errors for the JOB-light benchmark of our approach compared to the baselines. As we can see \system{} outperforms the best competitors in every percentile often by orders of magnitude. In the median it outperforms the best competitor Index Based Join Sampling (1.23 vs.~1.59). The advantage of the learned approach MCSN is that it outperforms traditional approaches by orders of magnitude for the higher percentiles and is thus more robust. Even for these outliers, \system{} provides additional robustness having a 95-th percentile for the q-errors of 3.16 vs.~143 (MCSN). The q-errors of both Postgres and random sampling are again significantly larger both for the medians and the higher percentiles. 
The estimation latencies for cardinalities using \system{} are currently in the order of $\mu$s to ms which suffices for complex join queries that can often run multiple $s$ on larger data sets. By using smaller RSPNs or an optimized implementation of SPNs such as \cite{sommer2018automatic}, the latencies could further be reduced.

\vspace{-1.5ex}\paragraph*{Generalization Capabilities} Especially for learned approaches the question of generalization is important, i.e. how well the models perform on previously unseen queries. 
For instance, by default the MCSN approach is only trained with queries up to three joins because otherwise the training data generation would be too expensive \cite{kipf2019learned}. 
Similarly in our approach, in the ensemble only few RSPNs with large joins occur because otherwise the training would also be too expensive. However, both approaches support cardinality estimates for unseen queries.

\begin{figure}
	\centering
	\includegraphics[width=0.99\linewidth]{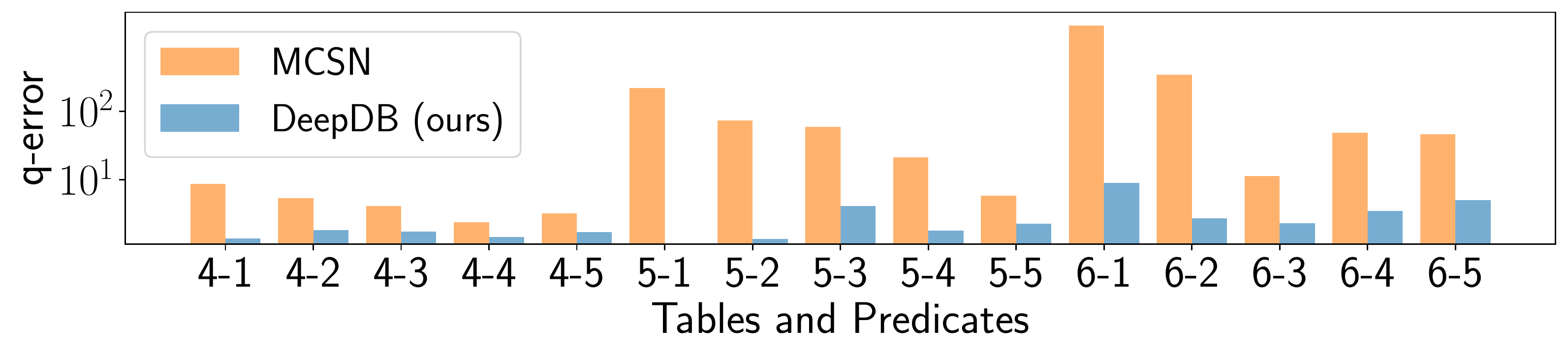}
	\vspace{-3.5ex}
	\caption{Median q-errors (logarithmic Scale) for different Join Sizes (4,5,6) and Number of Filter Predicates (1-5).}
	\vspace{-3.5ex}
	\label{fig:cardinalities_mcsn_join_size}
\end{figure}

To compare both learned approaches, we randomly generated queries for joins with four to six tables and one to five selection predicates for the IMDb data set. 
In Figure~\ref{fig:cardinalities_mcsn_join_size}, we plot the resulting median q-errors for both learned approaches: \system{} and MCSN \cite{kipf2019learned}. The median q-errors of \system{} are orders of magnitude lower for larger joins. Additionally, we can observe that for the MCSN approach the estimates tend to become less accurate for queries with fewer selection predicates. One possible explanation is that more tuples qualify for such queries and thus higher cardinalities have to be estimated. However, since there are at most three tables joined in the training data such higher cardinality values are most likely not predicted. We can conclude that using RSPNs leads to superior generalization capabilities.

\vspace{-1.5ex}\paragraph*{Updates}
\label{para:update-results}

In this experiment, we show that updated RSPN ensembles can precisely estimate cardinalities. To this end, we first learn the base RSPN ensemble on a certain share of the full IMDb data set (95\%, 90\%, 80\% and 60\%) and then update it using the remaining tuples. In a first experiment, the IMDb data set is randomly split while in the second experiment we learn the initial RSPNs on all movies up to a certain production year. Both experiments show that the q-error does not change significantly for the updated RSPN ensembles. Detailed results are given in Table~\ref{tab:updates}. We use zero as the budget factor to demonstrate that even base ensembles provide good estimates after updates. This is also the reason that the estimation errors slightly deviate from Table~\ref{fig:q_errors}.

\begin{table}[]
\begin{scriptsize}
    \centering
    \begin{tabular}{cccccc}
    \toprule
        Random &  0\% &   5\% &  10\% &  20\% &  40\%\\
        Split &        &        &        &        &       \\ \midrule
                          Median &  1.22 &  1.26 &  1.30 &  1.28 &  1.37 \\
                          90th &  3.45 &  3.04 &  2.94 &  3.15 &  3.60 \\
                          95th &  4.77 &  4.50 &  4.19 &  4.32 &  3.79 \\
        \toprule
        Temporal &  $<$ 2019 &  $<$ 2011 & $<$ 2009 & $<$ 2004 & $<$ 1991 \\
        Split & (0\%) & (4.7\%)  & (9.3\%)  & (19.7\%) & (40.1\%) \\ \midrule
                          Median & 1.22 & 1.28 &  1.31 &  1.34 & 1.41 \\
                          90th & 3.45 & 3.17 &  3.23 &  3.60 & 4.06 \\
                          95th & 4.77 & 4.30 &  3.83 &  4.07 & 4.35 \\
        \bottomrule
    \end{tabular}
    \vspace{-1.5ex}
    \caption{Estimation Errors for JOB-light after Updates for a random and temporal Split.}
    \label{tab:updates}
    \vspace{-4.5ex}
\end{scriptsize}
\end{table}

Since in the initial learning of the RSPN ensemble we learn the RSPN on a sample of the full outer join, the same sample rate has to be used for the updates, i.e. we only update the RSPN with a sample of all inserted tuples. Using a sampling rate of 1\%, we can handle up to 55,000 updates per second. 
The structure of the RSPN tree is not changed during updates, but only the parameters are updated according to the new tuples. However, in the experiments we could show that this does not impair the accuracy on a real-world data set.

The updateability is a clear advantage of \system{} compared to deep-learning based approaches for cardinality estimation \cite{kipf2019learned, sun2019an}. Since these model the problem end-to-end all training data queries would have to be run again on the database to gather the updated cardinalities.
 
\begin{figure}
	\centering
	\includegraphics[width=0.49\linewidth]{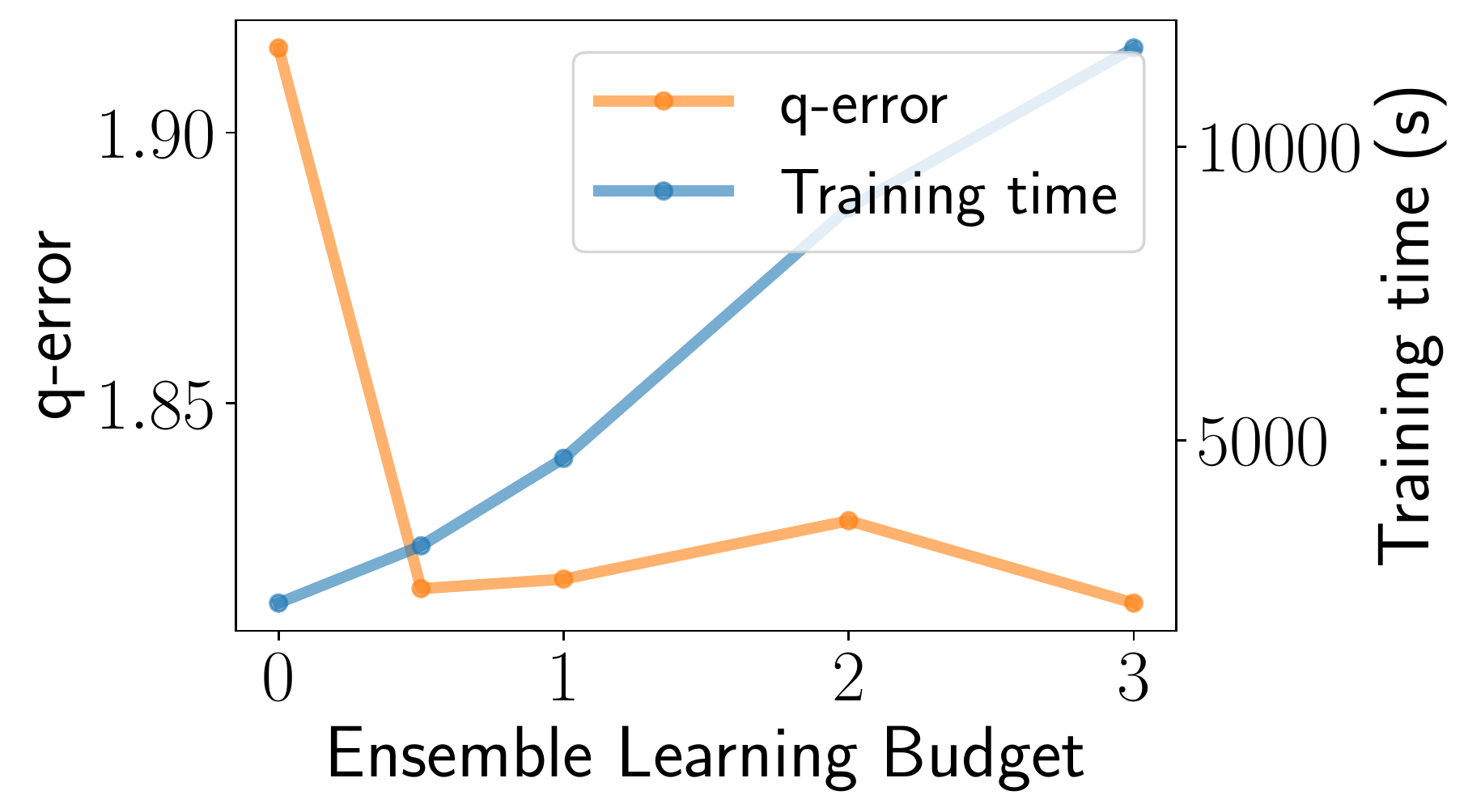}
	\includegraphics[width=0.49\linewidth]{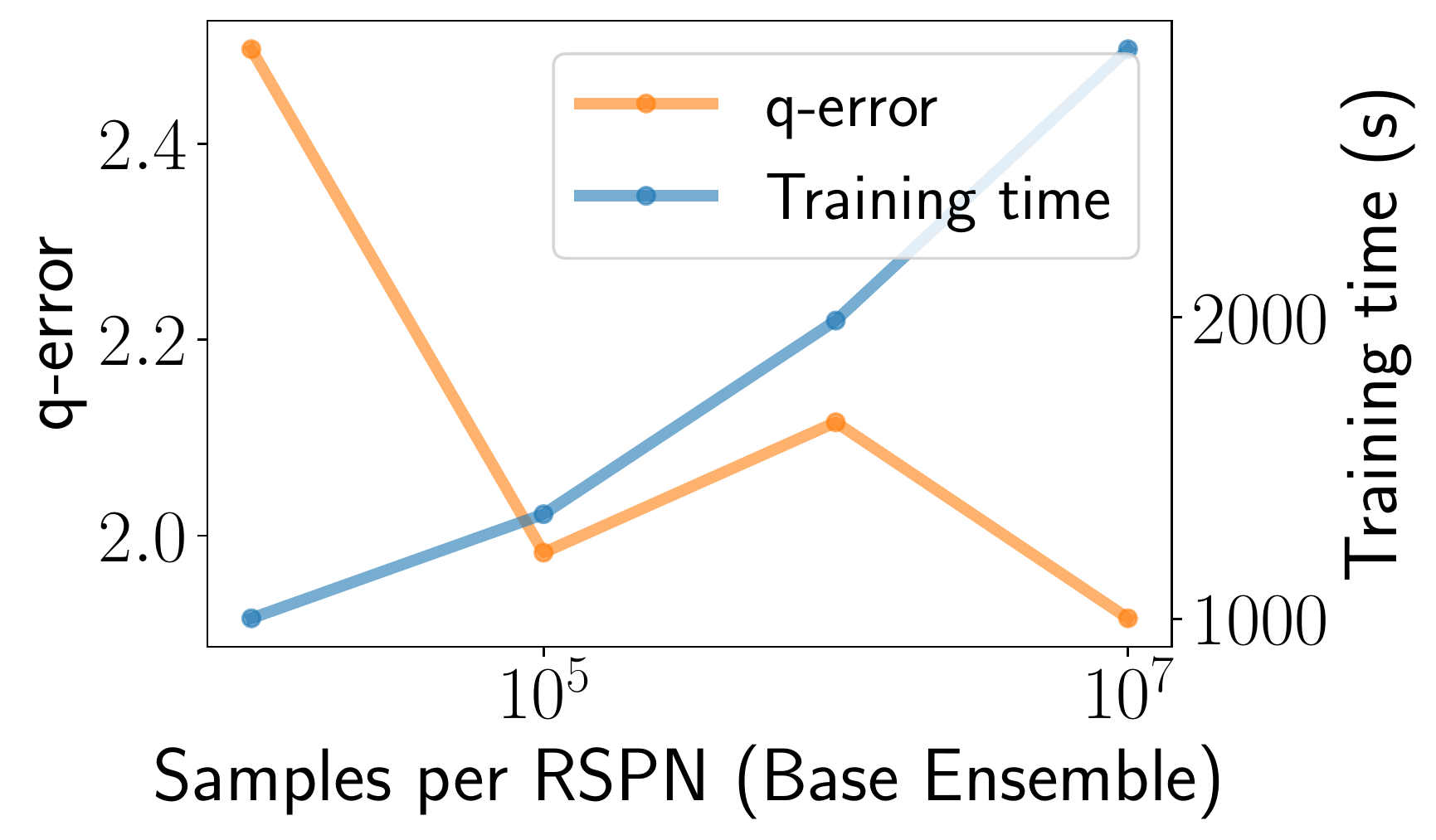}
	\vspace{-3.5ex}
	\caption{Q-errors and Training Time (in s) for varying Budget Factors and Sample Sizes.}
	\vspace{-3.5ex}
	\label{fig:param_exploration}
\end{figure}

\vspace{-1.5ex}\paragraph*{Parameter Exploration} Finally, in the last experiment we explore the tradeoff between ensemble training time and prediction quality of \system{}. We first vary the budget factor used in the ensemble selection between zero (i.e. learning only the base ensemble with one RSPN per join of two tables) and B=3 (i.e. the training of the larger RSPNs takes approximately three times longer than the base ensemble) while using $10^7$ samples per RSPN. We then use the resulting ensemble to evaluate 200 queries with three to six tables and one to five selection predicates. 
The resulting median q-errors are shown in Figure \ref{fig:param_exploration}. For higher budget factors the means are improving but already saturate at $B=0.5$. This is because there are no strong correlations in larger joins that have not already been captured in the base ensemble. 

We moreover evaluate the effect of the sampling to reduce the training time.
In this experiment we vary the sample size from 1000 to 10 million. We observe that while the training time increases, the higher we choose this parameter, the prediction quality improves (from 2.5 to 1.9 in the median). In summary, the training time can be significantly reduced if slight compromises in prediction quality are acceptable.
When minimization of training time is the more important objective we can also fall back and only learn RSPNs for all single tables and no joins at all. This reduces the ensemble training time to just five minutes. However, even this cheap strategy is still competitive. For JOB-light this ensemble has a median q-error of 1.98, a 90-th percentile of 5.32, a 95-th percentile of 8.54 and a maximum q-error of 186.53. Setting this in perspective to the baselines, this ensemble still outperforms state of the art for the higher percentiles and only Index Based Join Sampling is slightly superior in the median. This again proves the robustness of RSPNs. 

\subsection{Exp. 2: AQP}
\label{sec:aqp}

In this Section, we compare \system{} with state-of-the-art systems for AQP.

\vspace{-1.5ex}\paragraph*{Setup} We evaluated the approaches on both a synthetic data set and a real-world data set. As synthetic data set, we used the Star Schema Benchmark (SSB) \cite{o2009star} with a scale factor of $500$ with the standard queries (denoted by S1.1-S4.3). 
As the real-world data set, we used the Flights data set\footnote{https://www.kaggle.com/usdot/flight-delays} with queries ranging from selectivities between 5\% an 0.01\% covering a variety of group by attributes, \texttt{AVG}, \texttt{SUM} and \texttt{COUNT} queries (denoted by F1.1-F5.2). 
To scale the data set up to 1 billion records we used IDEBench \cite{eichmann2018idebench}. 

\vspace{-1.5ex}\paragraph*{Baselines} As baselines we used VerdictDB \cite{park2018verdict}, Wander Join \cite{li2016wander} and the Postgres \texttt{TABLESAMPLE} command (which uses random samples). VerdictDB is a middleware that can be used with any database system. It creates a stratified and a uniform sample for the fact tables to provide approximate queries. For VerdictDB, we used the default sample size (1\% of the full data set) for the Flights data set. For the SSB benchmark, this led to high query latencies and we thus decided to choose a sample size such that the query processing time was two seconds on average. 
Wander Join is a join sampling algorithm leveraging secondary indexes to generate join samples quickly. We set the time bound also to two seconds for a fair comparison and only evaluated this algorithm for data sets with joins. To this end, we created all secondary indexes for foreign key relationships and predicates on the dimension tables. For the \texttt{TABLESAMPLE} command we chose a sample percentage such that the queries take two seconds on average as well.

\begin{figure}
	\centering
	\includegraphics[width=\linewidth]{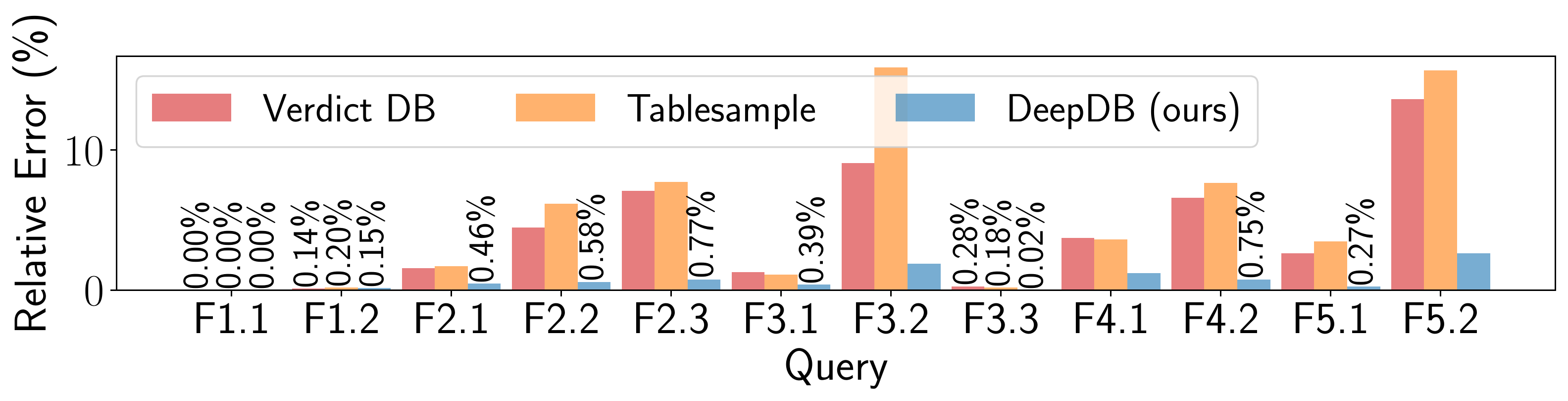}
	\includegraphics[width=\linewidth]{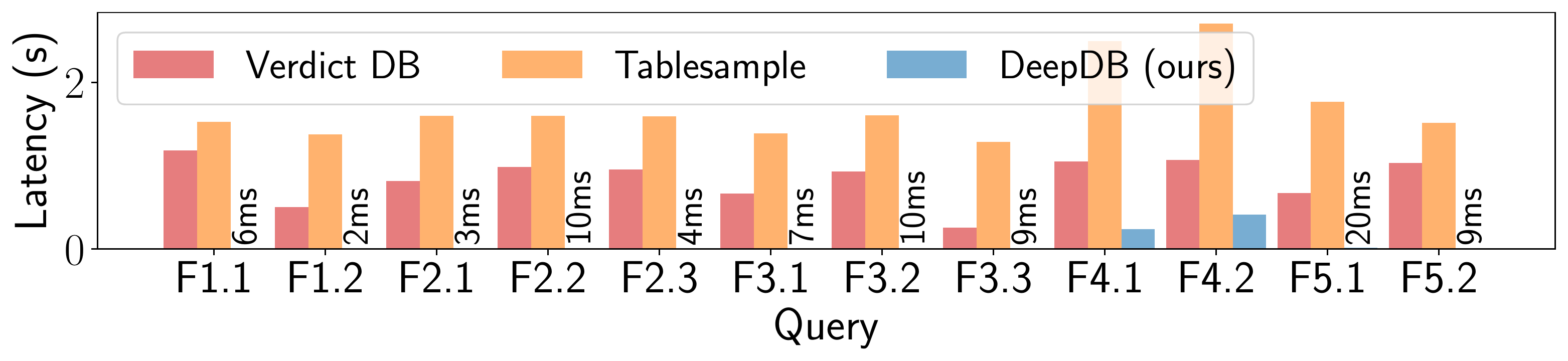}
	\vspace{-3.5ex}
	\caption{Average relative Error and Latencies for the Flights data set.}
	\label{fig:aqp_flights}
	\vspace{-3.5ex}
\end{figure}

\begin{figure*}
	\centering
	\includegraphics[width=0.8\linewidth]{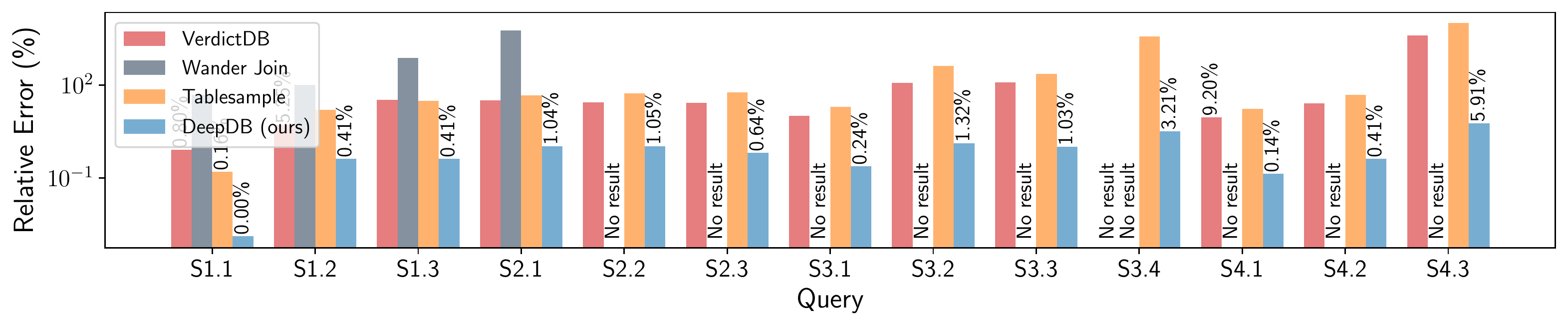}
	\vspace{-3.5ex}
	\caption{Average relative Error for SSB data set. Note the logarithmic Scale for the Errors.}
	\vspace{-3.5ex}
	\label{fig:aqp_ssb}
\end{figure*}

\vspace{-1.5ex}\paragraph*{Training Time}  

For \system{} the same hyperparameters were used as for the previous experiment (Exp. 1). The training took just 17 minutes for the SSB data set and 3 minutes for the Flights data set. The shorter training times compared to the IMDb data set are due to fewer cross-table correlations and hence fewer large join models in the ensemble. For VerdictDB, scrambles have to be created, i.e. uniform and stratified samples from the data set. This took 10 hours for the flights data set and 6 days for the SSB benchmark using the standard implementation.\footnote{https://docs.verdictdb.org/reference/pyverdict/} For wander join, secondary indexes had to be created also requiring several hours for the SSB data set.

\vspace{-1.5ex}\paragraph*{Accuracy and Latency}

For AQP two dimensions are of interest. First, the quality of the approximation quantified with the relative error. Second, the latency of the result is relevant when evaluating AQP systems. 
The relative error is defined as
$\frac{|a_{\mathit{true}}-a_{\mathit{predicted}}|}{a_{\mathit{true}}}$ where $a_{\mathit{true}}$ and $a_{\mathit{predicted}}$ are the true and predicted aggregate function, respectively. 
If the query is a group by query, several aggregates have to be computed. In this case, the relative error is averaged over all groups. 
The results for the Flights data set are given in Figure \ref{fig:aqp_flights}.

For the Flights data set, we can observe that \system{} always has the lowest average relative error. This is often the case for queries with lower selectivities where sample-based approaches have few tuples that satisfy the selection predicates and thus the approximations are very inaccurate. In contrast, \system{} does not rely on samples but models the data distribution and leverages the learned representation to provide estimates. For instance, for query 11 with a selectivity of 0.5\% VerdictDB and the \texttt{TABLESAMPLE} strategy have an average relative error of 15.6\% and 13.6\%, respectively. In contrast, the average relative error of \system{} is just 2.6\%.

Moreover, the latencies for both \texttt{TABLESAMPLE} and VerdictDB are between one and two seconds on average. In contrast, \system{} does not rely on sampling but on evaluating the RSPNs. This is significantly faster resulting in a maximum latency of 31ms. This even holds true for queries with several groups where more expectations have to be computed (at least one additional per different group).

The higher accuracies of \system{} are even more severe for the SSB benchmark. The queries have even lower selectivities between 3.42\% and 0.0075\% for queries 1 to 12 and 0.00007\% for the very last query. This results in very inaccurate predictions of the sample-based approaches. Here, the average relative errors are orders of magnitude lower for \system{} always being less than 6\%. In contrast, VerdictDB, Wander Join and the \texttt{TABLESAMPLE} approach often have average relative errors larger than 100\%. Moreover, for some queries no estimate can be given at all because no samples are drawn that satisfy the filter predicates. However, while the other approaches take two seconds to provide an estimate, \system{} requires no more than 293ms in the worst case. In general the latencies for \system{} are lower for queries with fewer groups because less expectations have to be computed.

\vspace{-1.5ex}\paragraph*{Confidence Intervals} In this experiment, we evaluate how accurate the confidence intervals predicted by \system{} are. To this end, we measure the relative confidence interval length defined as: $\frac{a_{\mathit{predicted}}-a_{\mathit{lower}}}{a_{\mathit{predicted}}}$, where $a_{\mathit{predicted}}$ is the prediction and $a_{\mathit{lower}}$ is the lower bound of the confidence interval. 

\begin{figure}
	\centering
	\includegraphics[width=\linewidth]{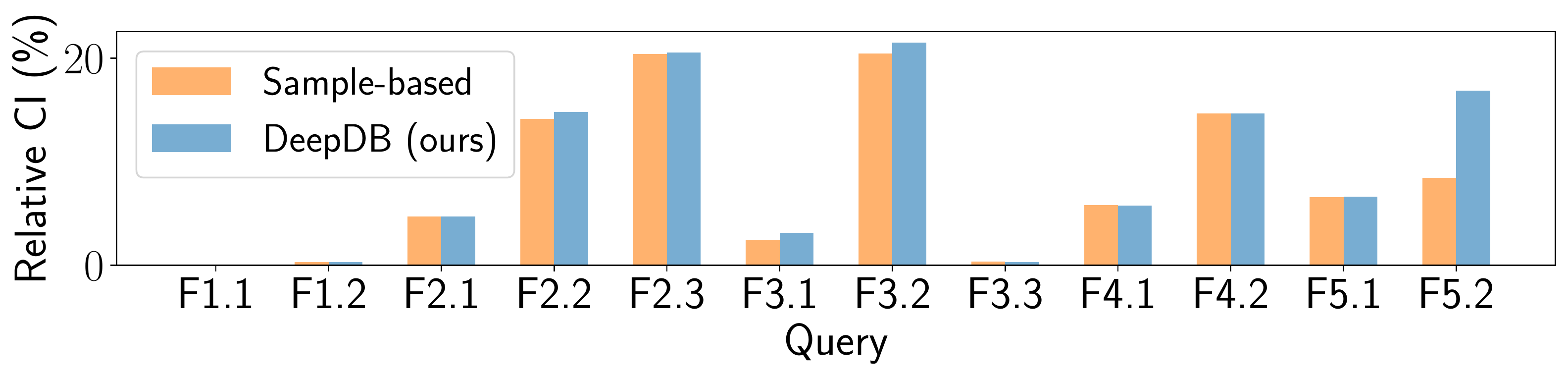}
	\includegraphics[width=\linewidth]{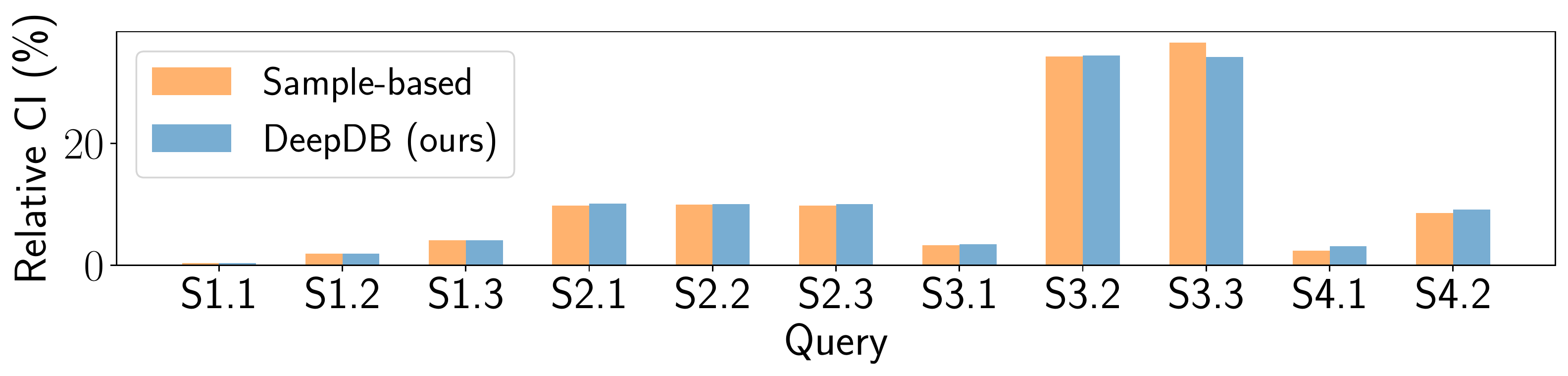}
	\vspace{-3.5ex}
	\caption{True and predicted relative length of the Confidence Intervals.}
	\label{fig:aqp_ci}
	\vspace{-3.5ex}
\end{figure}

This relative confidence interval length is compared to the confidence interval of a sample-based approach. For this we draw 10 million samples (as many samples as our models use for learning in this experiment) and compute estimates for the average, count and sum aggregates. We then compute the confidence intervals of these estimates using standard statistical methods. For \texttt{COUNT} aggregates, the estimator is simply a binomial variable with parameters $n=n_{\mathit{samples}}$ and $p=a_{\mathit{predicted}}/n_{\mathit{samples}}$ for which we can compute a confidence interval. For \texttt{AVG} queries we exploit the central limit theorem stating that the estimator is normally distributed. We then compute the standard deviation on the sample and derive the confidence interval for a normal variable having this standard deviation and the mean of our estimate. For \texttt{SUM} queries we model the estimator as a product of both estimators. The resulting confidence interval lengths can be seen as ground truth and are compared to the confidence intervals of our system in Figure \ref{fig:aqp_ci}. Note that we excluded queries where less than 10 samples fulfilled the filter predicates. In these cases the estimation of a standard deviation has itself a too high variance.

In all cases, the confidence intervals of \system{} are very good approximations of the true confidence intervals. The only exception is query F5.2 for the Flights data set which is a difference of two \texttt{SUM} aggregates. In this case, assumption (i) of Section \ref{sec:confidence_intervals} does not hold: the probabilities and expectation estimates cannot be considered independent. This is the case because both \texttt{SUM} aggregates contain correlated attributes and thus the confidence intervals are overestimated. However, note that in the special case of the difference of two sum aggregates the AQP estimates are still very precise as shown in Figure \ref{fig:aqp_flights} for the same query F5.2. Only the confidence interval is overestimated. Such cases can easily be identified and only occur when arithmetic expressions of several aggregates should be estimated.

\vspace{-1.5ex}\paragraph*{Other ML-based Approaches} The only learned approach for AQP that was recently published is DBEst \cite{ma2019dbest}. Other approaches like \cite{thirumuruganathan2019approximate} cannot provide estimates for joins and are thus similarly excluded. 
DBEst creates density and regression models for popular queries. They can be reused if only where conditions on numeric attributes or ordinal categorical attributes are changed. But if an unseen new query arrives and there is no model available we have to create a biased sample fulfilling the non-ordinal conditions on categorical columns. 

\begin{figure}
	\centering
	\includegraphics[width=0.9 \linewidth]{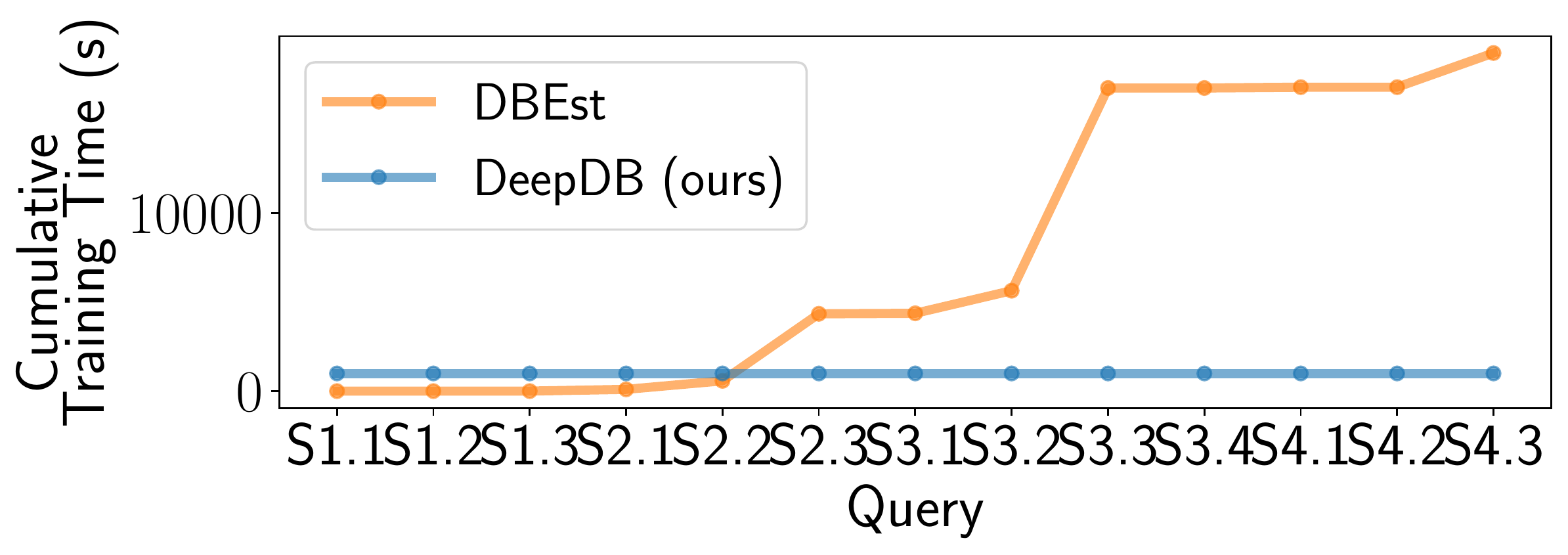}
	\vspace{-1.5ex}
	\caption{Cumulative Time to create DBEst and \system{} models for SSB Queries.}
	\label{fig:dbest_creation_time}
	\vspace{-1.5ex}
\end{figure}

Depending on the selectivity, this comes at a cost. Afterwards, the density and regression models on the sample have to be learned. In contrast, in our approach, we learn an RSPN ensemble once and can provide estimates for arbitrary queries immediately. In Figure \ref{fig:dbest_creation_time}, we thus compare the cumulative training time including sampling and data preparation times of DBEst and \system{} for SSB. As we can see, for query S1.2 and S1.3 the model of query S1.1 can be reused and thus the cumulative training time does not increase. In contrast, for some selective queries like S3.3 the biased sampling and training takes very long ($>$ 3 hours). For \system{} the ensemble has to be trained just once and any query can be answered ad-hoc. 

\subsection{Exp. 3: Machine Learning}

\begin{figure}
	\centering
	\includegraphics[width=0.9\linewidth]{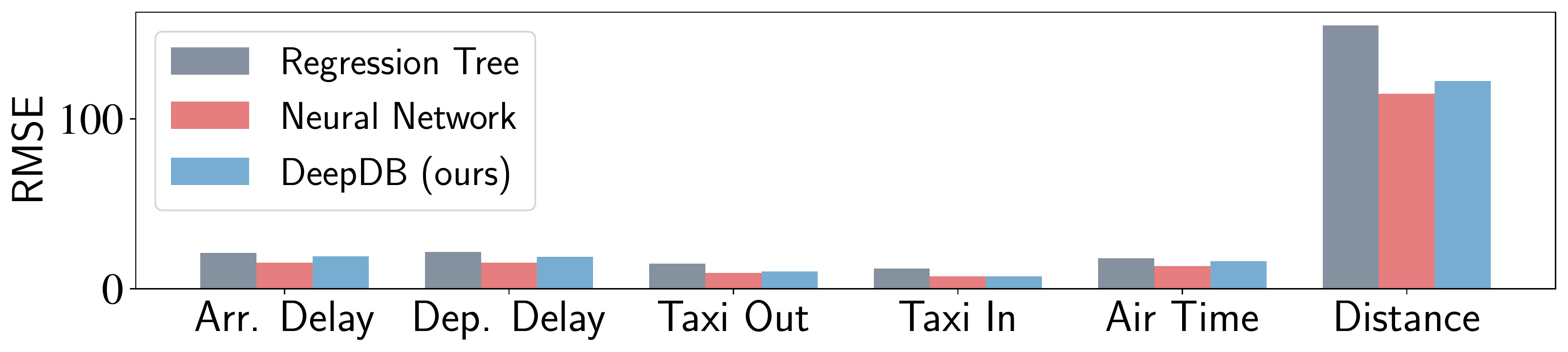}
	\includegraphics[width=0.9\linewidth]{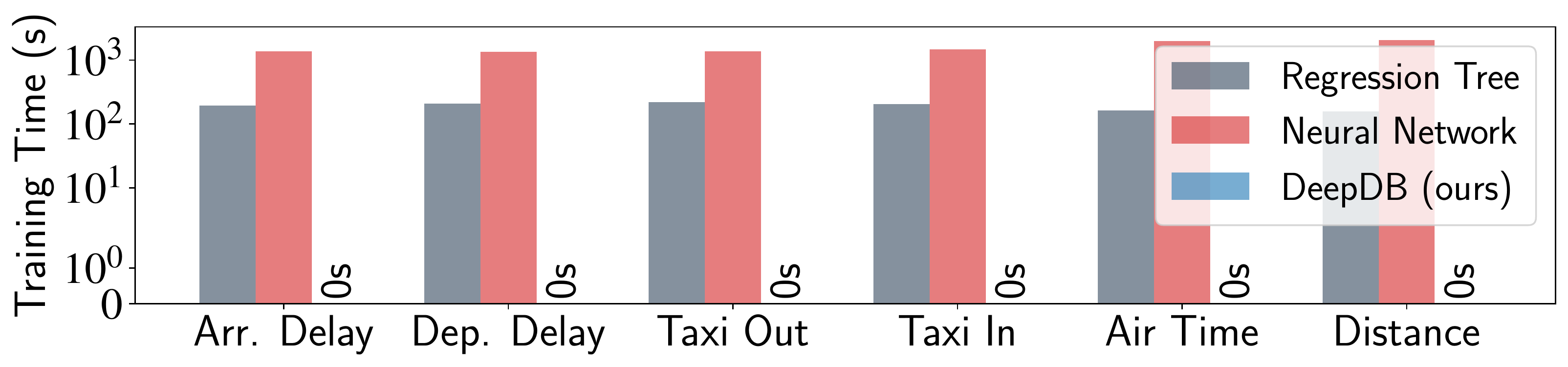}
	\vspace{-1.5ex}
	\caption{Estimation Errors and Training Times for Regression Tasks.}
	\label{fig:regression}
	\vspace{-3.5ex}
\end{figure}

In this experiment we show that RSPNs are competitive ML regression models. We first predict all different numeric attributes for the Flights data set using all other columns as features with the same RSPN we used for the AQP queries. As baselines we trained standard ML models on the same training data to solve the same prediction task and compare the training time and Root Mean Squared Error (RMSE) on the test set in Figure \ref{fig:regression}. The advantage of \system{} is that no additional training is required to execute the regression task while the RMSE is comparable to standard models. Consequently, using RSPNs we obtain a free classification and regression model for any combination of features.

 \balance{}
\section{Related Work}
\label{sec:related}

Before concluding, we discuss further related work on using ML for cardinality estimation and AQP and SPNs.

\vspace{-1.5ex}\paragraph*{Learned Cardinality Estimation} The problem of selectivity estimation for single tables is a special case of cardinality estimation. There is a large body of work applying different ML approaches including probabilistic graphical models \cite{tzoumas2013efficiently, getoor2001selectivity, getoor2011learning}, neural networks \cite{lakshmi1998selectivity, liu2015cardinality} and specialized deep learning density models \cite{shohedul2019multiattribute} to this problem. Recently, Dutt et al. \cite{selectivity2019lightweight} suggested using lightweight tree-based models in combination with log-transformed labels leading to superior predictions.

The first works applying ML to cardinality estimation including joins used simple regression models \cite{akdere2012learning, malik07ablackbox}. More recently, Deep Learning was specifically proposed to solve cardinality estimation end-to-end \cite{kipf2019learned, sun2019an}.
Woltmann et al. \cite{woltmann2019cardinality} also separate the problem of cardinality estimation on a large schema. To this end, deep learning models similar to \cite{kipf2019learned} are learned for certain schema sub-parts. However, two models for schema sub-parts cannot be combined to provide estimates for a larger join. Other techniques exploit learned models for overlapping subgraph templates for recurring cloud workloads \cite{wu2018towards}.
All these models need a workload to be executed and used as training data which is different from our data-driven approach.

\vspace{-1.5ex}\paragraph*{Learned AQP} 
Early work \cite{shanmugasundaram1999compressed} suggests to approximate OLAP cubes by mixture models based on found clusters in the data. Though greatly reducing the required storage, the approximation errors are relatively high. \emph{FunctionDB} \cite{thiagarajan2008querying} constructs piecewise linear functions as approximation. In contrast to \system{}, only continuous variables are supported. \emph{DBEst} \cite{ma2019dbest} builds models for popular queries and thus samples do not have to be kept. However, in contrast to \system{} only those popular queries and no ad-hoc queries are supported. Park et al. suggested \emph{Database Learning} \cite{park2019database} which builds a model from query results that is leveraged to provide approximate results for future queries. In contrast, \system{} is data-driven and does not require past query results. Moreover, specialized generative models were suggested to draw samples for AQP \cite{thirumuruganathan2019approximate}. However, this technique does not work for joins.

\vspace{-1.5ex}\paragraph*{SPNs} Sum Product Networks \cite{domingos2011spn, molina2017poisson, molina2017mixed} have recently gained attention because these graphical models allow an efficient inference process. Furthermore, our update process can be seen as an orthogonal approach to online learning for SPNs \cite{kalra2018online}. In contrast to incremental learning schemes \cite{lee2013online} for SPNs, we do not change the structure if new tuples are inserted for performance reasons.

 \section{Conclusion and Future work}
\label{sec:summary}

In this work we have proposed \system{} which is a data-driven approach for learned database components. We have shown that our approach is general and can be used to support various tasks including cardinality estimation and approximate query processing. Our experiments demonstrate that \system{} outperforms both traditional and learned state-of-the-art techniques often by orders of magnitude. In addition, we leveraged the same approach to support ML tasks on the data set with accuracies competitive with neural networks while not requiring any additional training time.

We believe our data-driven approach for learning can also be exploited to improve other database internals. For instance, it has already been shown that column correlations can be exploited to improve indexing \cite{wu2019designing}. In addition, SPNs naturally provide a notion of correlated clusters that can also be used for suggesting using interesting patterns in data exploration. Finally, we believe that it is an interesting avenue of future work to combine data-driven and workload-driven approaches to combine the best of both worlds.

\newpage
\begin{scriptsize}
\bibliographystyle{abbrv}
\bibliography{deepdb}  

\begin{thebibliography}{10}

\bibitem{akdere2012learning}
M.~Akdere, U.~{\c{C}}etintemel, M.~Riondato, E.~Upfal, and S.~B. Zdonik.
\newblock Learning-based query performance modeling and prediction.
\newblock In {\em 2012 IEEE 28th International Conference on Data Engineering},
  pages 390--401. IEEE, 2012.

\bibitem{selectivity2019lightweight}
A.~Dutt, C.~Wang, A.~Nazi, S.~Kandula, V.~R. Narasayya, and S.~Chaudhuri.
\newblock Selectivity estimation for range predicates using lightweight models.
\newblock {\em {PVLDB}}, 12(9):1044--1057, 2019.

\bibitem{eichmann2018idebench}
P.~Eichmann, C.~Binnig, T.~Kraska, and E.~Zgraggen.
\newblock Idebench: A benchmark for interactive data exploration, 2018.

\bibitem{getoor2011learning}
L.~Getoor and L.~Mihalkova.
\newblock Learning statistical models from relational data.
\newblock In {\em Proceedings of the 2011 ACM SIGMOD International Conference
  on Management of Data}, SIGMOD '11, pages 1195--1198, New York, NY, USA,
  2011. ACM.

\bibitem{getoor2001selectivity}
L.~Getoor, B.~Taskar, and D.~Koller.
\newblock Selectivity estimation using probabilistic models.
\newblock In {\em Proceedings of the 2001 ACM SIGMOD International Conference
  on Management of Data}, SIGMOD '01, pages 461--472, New York, NY, USA, 2001.
  ACM.

\bibitem{shohedul2019multiattribute}
S.~Hasan, S.~Thirumuruganathan, J.~Augustine, N.~Koudas, and G.~Das.
\newblock Multi-attribute selectivity estimation using deep learning.
\newblock {\em CoRR}, abs/1903.09999, 2019.

\bibitem{kalra2018online}
A.~Kalra, A.~Rashwan, W.-S. Hsu, P.~Poupart, P.~Doshi, and G.~Trimponias.
\newblock Online structure learning for feed-forward and recurrent sum-product
  networks.
\newblock In {\em Advances in Neural Information Processing Systems}, pages
  6944--6954, 2018.

\bibitem{kipf2019learned}
A.~Kipf, T.~Kipf, B.~Radke, V.~Leis, P.~A. Boncz, and A.~Kemper.
\newblock Learned cardinalities: Estimating correlated joins with deep
  learning.
\newblock In {\em {CIDR} 2019, 9th Biennial Conference on Innovative Data
  Systems Research, Asilomar, CA, USA, January 13-16, 2019, Online
  Proceedings}, 2019.

\bibitem{kraska2018thecase}
T.~Kraska, A.~Beutel, E.~H. Chi, J.~Dean, and N.~Polyzotis.
\newblock The case for learned index structures.
\newblock In {\em Proceedings of the 2018 International Conference on
  Management of Data}, SIGMOD '18, pages 489--504, New York, NY, USA, 2018.
  ACM.

\bibitem{lakshmi1998selectivity}
M.~S. Lakshmi and S.~Zhou.
\newblock Selectivity estimation in extensible databases - a neural network
  approach.
\newblock In {\em Proceedings of the 24rd International Conference on Very
  Large Data Bases}, VLDB '98, pages 623--627, San Francisco, CA, USA, 1998.
  Morgan Kaufmann Publishers Inc.

\bibitem{lee2013online}
S.-W. Lee, M.-O. Heo, and B.-T. Zhang.
\newblock Online incremental structure learning of sum--product networks.
\newblock In {\em International Conference on Neural Information Processing},
  pages 220--227. Springer, 2013.

\bibitem{leis2015how}
V.~Leis, A.~Gubichev, A.~Mirchev, P.~Boncz, A.~Kemper, and T.~Neumann.
\newblock How good are query optimizers, really?
\newblock {\em Proc. VLDB Endow.}, 9(3):204--215, Nov. 2015.

\bibitem{leis2017ibjs}
V.~Leis, B.~Radke, A.~Gubichev, A.~Kemper, and T.~Neumann.
\newblock Cardinality estimation done right: Index-based join sampling.
\newblock In {\em {CIDR} 2017, 8th Biennial Conference on Innovative Data
  Systems Research, Chaminade, CA, USA, January 8-11, 2017, Online
  Proceedings}, 2017.

\bibitem{li2016wander}
F.~Li, B.~Wu, K.~Yi, and Z.~Zhao.
\newblock Wander join: Online aggregation via random walks.
\newblock In {\em Proceedings of the 2016 International Conference on
  Management of Data}, SIGMOD '16, pages 615--629, New York, NY, USA, 2016.
  ACM.

\bibitem{liu2015cardinality}
H.~Liu, M.~Xu, Z.~Yu, V.~Corvinelli, and C.~Zuzarte.
\newblock Cardinality estimation using neural networks.
\newblock In {\em Proceedings of the 25th Annual International Conference on
  Computer Science and Software Engineering}, CASCON '15, pages 53--59,
  Riverton, NJ, USA, 2015. IBM Corp.

\bibitem{lopez2013randomized}
D.~Lopez-Paz, P.~Hennig, and B.~Sch{\"o}lkopf.
\newblock The randomized dependence coefficient.
\newblock In {\em Advances in neural information processing systems}, pages
  1--9, 2013.

\bibitem{DBLP:conf/sigmod/MaT19}
Q.~Ma and P.~Triantafillou.
\newblock Dbest: Revisiting approximate query processing engines with machine
  learning models.
\newblock In {\em Proceedings of the 2019 International Conference on
  Management of Data, {SIGMOD} Conference 2019, Amsterdam, The Netherlands,
  June 30 - July 5, 2019.}, pages 1553--1570, 2019.

\bibitem{ma2019dbest}
Q.~Ma and P.~Triantafillou.
\newblock Dbest: Revisiting approximate query processing engines with machine
  learning models.
\newblock In {\em Proceedings of the 2019 International Conference on
  Management of Data}, SIGMOD '19, pages 1553--1570, New York, NY, USA, 2019.
  ACM.

\bibitem{malik07ablackbox}
T.~Malik, R.~Burns, and N.~Chawla.
\newblock A black-box approach to query cardinality estimation.
\newblock In {\em CIDR}, 2007.

\bibitem{DBLP:journals/corr/abs-1904-03711}
R.~Marcus, P.~Negi, H.~Mao, C.~Zhang, M.~Alizadeh, T.~Kraska, O.~Papaemmanouil,
  and N.~Tatbul.
\newblock Neo: {A} learned query optimizer.
\newblock {\em CoRR}, abs/1904.03711, 2019.

\bibitem{molina2017poisson}
A.~Molina, S.~Natarajan, and K.~Kersting.
\newblock {Poisson Sum-Product Networks: A Deep Architecture for Tractable
  Multivariate Poisson Distributions}.
\newblock 2017.

\bibitem{molina2017mixed}
A.~Molina, A.~Vergari, N.~D. Mauro, S.~Natarajan, F.~Esposito, and K.~Kersting.
\newblock {Mixed Sum-Product Networks: A Deep Architecture for Hybrid Domains}.
\newblock In {\em AAAI}, 2018.

\bibitem{molina2019spflow}
A.~Molina, A.~Vergari, K.~Stelzner, R.~Peharz, P.~Subramani, N.~D. Mauro,
  P.~Poupart, and K.~Kersting.
\newblock Spflow: An easy and extensible library for deep probabilistic
  learning using sum-product networks, 2019.

\bibitem{nath2015learning}
A.~Nath and P.~Domingos.
\newblock Learning relational sum-product networks.
\newblock In {\em Proceedings of the Twenty-Ninth AAAI Conference on Artificial
  Intelligence}, pages 2878--2886. AAAI Press, 2015.

\bibitem{o2009star}
P.~O’Neil, E.~O’Neil, X.~Chen, and S.~Revilak.
\newblock The star schema benchmark and augmented fact table indexing.
\newblock In {\em Technology Conference on Performance Evaluation and
  Benchmarking}, pages 237--252. Springer, 2009.

\bibitem{park2018verdict}
Y.~Park, B.~Mozafari, J.~Sorenson, and J.~Wang.
\newblock Verdictdb: Universalizing approximate query processing.
\newblock In {\em Proceedings of the 2018 International Conference on
  Management of Data}, SIGMOD '18, pages 1461--1476, New York, NY, USA, 2018.
  ACM.

\bibitem{park2019database}
Y.~Park, A.~S. Tajik, M.~Cafarella, and B.~Mozafari.
\newblock Database learning: Toward a database that becomes smarter every time.
\newblock In {\em Proceedings of the 2017 ACM International Conference on
  Management of Data}, SIGMOD '17, pages 587--602, New York, NY, USA, 2017.
  ACM.

\bibitem{domingos2011spn}
H.~Poon and P.~Domingos.
\newblock {Sum-product networks: A New Deep Architecture}.
\newblock In {\em 2011 {IEEE} {International} {Conference} on {Computer}
  {Vision} {Workshops}}, pages 689--690, November 2011.

\bibitem{shanmugasundaram1999compressed}
J.~Shanmugasundaram, U.~Fayyad, P.~S. Bradley, et~al.
\newblock Compressed data cubes for olap aggregate query approximation on
  continuous dimensions.
\newblock 1999.

\bibitem{DBLP:conf/sigmod/ShengTZP19}
Y.~Sheng, A.~Tomasic, T.~Zhang, and A.~Pavlo.
\newblock Scheduling {OLTP} transactions via learned abort prediction.
\newblock In {\em Proceedings of the Second International Workshop on
  Exploiting Artificial Intelligence Techniques for Data Management,
  aiDM@SIGMOD 2019, Amsterdam, The Netherlands, July 5, 2019}, pages 1:1--1:8,
  2019.

\bibitem{sommer2018automatic}
L.~{Sommer}, J.~{Oppermann}, A.~{Molina}, C.~{Binnig}, K.~{Kersting}, and
  A.~{Koch}.
\newblock Automatic mapping of the sum-product network inference problem to
  fpga-based accelerators.
\newblock In {\em 2018 IEEE 36th International Conference on Computer Design
  (ICCD)}, pages 350--357, 2018.

\bibitem{sun2019an}
J.~Sun and G.~Li.
\newblock An end-to-end learning-based cost estimator.
\newblock {\em CoRR}, abs/1906.02560, 2019.

\bibitem{thiagarajan2008querying}
A.~Thiagarajan and S.~Madden.
\newblock Querying continuous functions in a database system.
\newblock In {\em Proceedings of the 2008 ACM SIGMOD international conference
  on Management of data}, pages 791--804. ACM, 2008.

\bibitem{thirumuruganathan2019approximate}
S.~Thirumuruganathan, S.~Hasan, N.~Koudas, and G.~Das.
\newblock Approximate query processing using deep generative models.
\newblock {\em CoRR}, abs/1903.10000, 2019.

\bibitem{tzoumas2013efficiently}
K.~Tzoumas, A.~Deshpande, and C.~S. Jensen.
\newblock Efficiently adapting graphical models for selectivity estimation.
\newblock {\em The VLDB Journal}, 22(1):3--27, Feb. 2013.

\bibitem{woltmann2019cardinality}
L.~Woltmann, C.~Hartmann, M.~Thiele, D.~Habich, and W.~Lehner.
\newblock Cardinality estimation with local deep learning models.
\newblock In {\em Proceedings of the Second International Workshop on
  Exploiting Artificial Intelligence Techniques for Data Management}, aiDM '19,
  pages 5:1--5:8, New York, NY, USA, 2019. ACM.

\bibitem{wu2018towards}
C.~Wu, A.~Jindal, S.~Amizadeh, H.~Patel, W.~Le, S.~Qiao, and S.~Rao.
\newblock Towards a learning optimizer for shared clouds.
\newblock {\em Proc. VLDB Endow.}, 12(3):210--222, Nov. 2018.

\bibitem{wu2019designing}
Y.~Wu, J.~Yu, Y.~Tian, R.~Sidle, and R.~Barber.
\newblock Designing succinct secondary indexing mechanism by exploiting column
  correlations.
\newblock In {\em Proceedings of the 2019 International Conference on
  Management of Data}, SIGMOD '19, pages 1223--1240, New York, NY, USA, 2019.
  ACM.

\end{thebibliography}
\end{scriptsize}

\end{document}